\documentclass{jfm}

\usepackage{graphicx}
\usepackage{epstopdf, epsfig}
\usepackage{natbib}
\usepackage{upmath}
\usepackage{color}
\usepackage{subfig}
\usepackage{booktabs}
\usepackage{amssymb}%
\usepackage{amsmath}
\usepackage{multirow}
\usepackage{pgfplots}
\let\le=\leqslant  \let\leq=\leqslant
  
\usepackage{amsbsy}
\allowdisplaybreaks

\title[Spectrograms of ship wakes]{Spectrograms of ship wakes: \\ identifying linear and nonlinear wave signals}

\author[R. Pethiyagoda, S. W. McCue and T. J. Moroney]%
{\small Ravindra Pethiyagoda$^1$,\ns
Scott W. McCue$^1$\thanks{Email address for correspondence: scott.mccue@qut.edu.au}
and Timothy J. Moroney$^1$}

\affiliation{\small $^1$School of Mathematical Sciences, Queensland University of Technology, Brisbane QLD 4001, Australia}

\pubyear{}
\volume{}
\pagerange{}
\date{\today}
\begin{document}

\maketitle

\begin{abstract}
A spectrogram is a useful way of using short-time discrete Fourier transforms to visualise surface height measurements taken of ship wakes in real world conditions.   For a steadily moving ship that leaves behind small-amplitude waves, the spectrogram is known to have two clear linear components, a sliding-frequency mode caused by the divergent waves and a constant-frequency mode for the transverse waves.  However, recent observations of high speed ferry data have identified additional components of the spectrograms that are not yet explained.  We use computer simulations of linear and nonlinear ship wave patterns and apply time-frequency analysis to generate spectrograms for an idealised ship.  We clarify the role of the linear dispersion relation and ship speed on the two linear components.  We use a simple weakly nonlinear theory to identify higher order effects in a spectrogram and, while the high speed ferry data is very noisy, we propose that certain additional features in the experimental data are caused by nonlinearity.  Finally, we provide a possible explanation for a further discrepancy between the high speed ferry spectrograms and linear theory by accounting for ship acceleration.

\end{abstract}


\section{Introduction}

A { useful} method for observing and measuring ship wakes is to employ an echo sounder to record the water height over time as a ship passes nearby.  The resulting output signal corresponds to the cross-section of the ship wake taken in the direction of travel \citep{torsvik15a}.  The surface elevation at the echo sounder can be visualised as a spectrogram through the use of many short-time discrete Fourier transforms.  In this paper, we aim to identify and explain features of spectrograms of ship wakes, concentrating on the differing effects that linearity and nonlinearity have on the wave time-frequency signal.

The study of ship waves has been of great academic interest for over a century \citep{darrigol03}. From a mathematical perspective, a popular approach is to set up a potential flow model, with the effects of the ship approximated by a steadily moving pressure distribution acting on the surface of the water.  By linearising the dynamic and kinematic boundary conditions on the free surface, one can write down exact solutions for the wave pattern using a Fourier transform.  In this way, theoretical studies provide insight into how the speed of a ship affects the distinguishing features such as the divergent and transverse waves \citep{peters49,ursell60,chung91} and the wake angle \citep{rabaud13,darmon14,noblesse14,pethiyagoda15}.
Similarly, linear water wave theory has been used extensively to approximate the drag force associated with a ship wake \citep{michell98,havelock32,noblesse81}.  For the fully nonlinear versions of these problems, direct analysis is much more difficult \citep{soomere07}.  Thus, studying the effects of nonlinearity on a ship wake is normally framed as a computational challenge \citep{forbes89,parau02,parau07,pethiyagoda14a}.

In the real world, analysis of ship wave data is limited due to the difficulty of accurately capturing the surface height outside of a towing tank. One method of observing ship wakes is with satellite photography \citep{munk86,rabaud13} or radar \citep{milgram88,reed02}. Unfortunately, satellite photography requires adequate lighting conditions to highlight the desired wake components \citep{munk86}. Even with sufficient clarity, a photograph can not provide quantitative data on the surface elevation, leaving only a few viable measurements to be performed on the ship wake, such as the wake angle \citep{rabaud13}. Radar can be hampered by backscattering that leads to a bright narrow V-pattern \citep{milgram88} caused by Bragg-resonant ship waves \citep{reed02} and not related to the wake angles observed by \citet{rabaud13}.  Another method of observing ship wakes is to use an echo sounder and to visualise the frequencies via a spectrogram, as mentioned above.  We will focus on spectrograms in this paper.

Spectrograms are popular in many fields and, for example, have been used in signal processing for decades \citep{cohen89} to decompose signals into wave components of different frequency. Even though \citet{tuck71} determined the theoretical recovered wave frequencies for a sensor travelling over a far-field ship wake, spectrograms are a relatively new tool in the study of water waves, originally used by \citet{wyatt88} to analyse ship wakes with a sensor moving perpendicular to the direction of the ship. {The work of \citet{wyatt88} was closely followed by \citet{brown89} who performed experiments with a stationary sensor and constructed low resolution spectrograms using experimental data.  There has been a resurgence of interest in the use of spectrograms for analysing ship waves \citep{benassai15,didenkulova13,sheremet13,torsvik15a,torsvik15b}, for which a stationary sensor is used to measure the wake of ships.}
For much of this recent work, the primary focus of the research has been on calculating the energy contained in a given wake and the effect that the propagating wake wash will have when it interacts with the coastal zone \citep{benassai15,didenkulova13,torsvik15a}.  This work is important because regulators need to balance the protection of the coastal environment (both natural and built) against the need for efficient shipping systems.

\begin{figure}
\centering
\includegraphics[width=.8\linewidth]{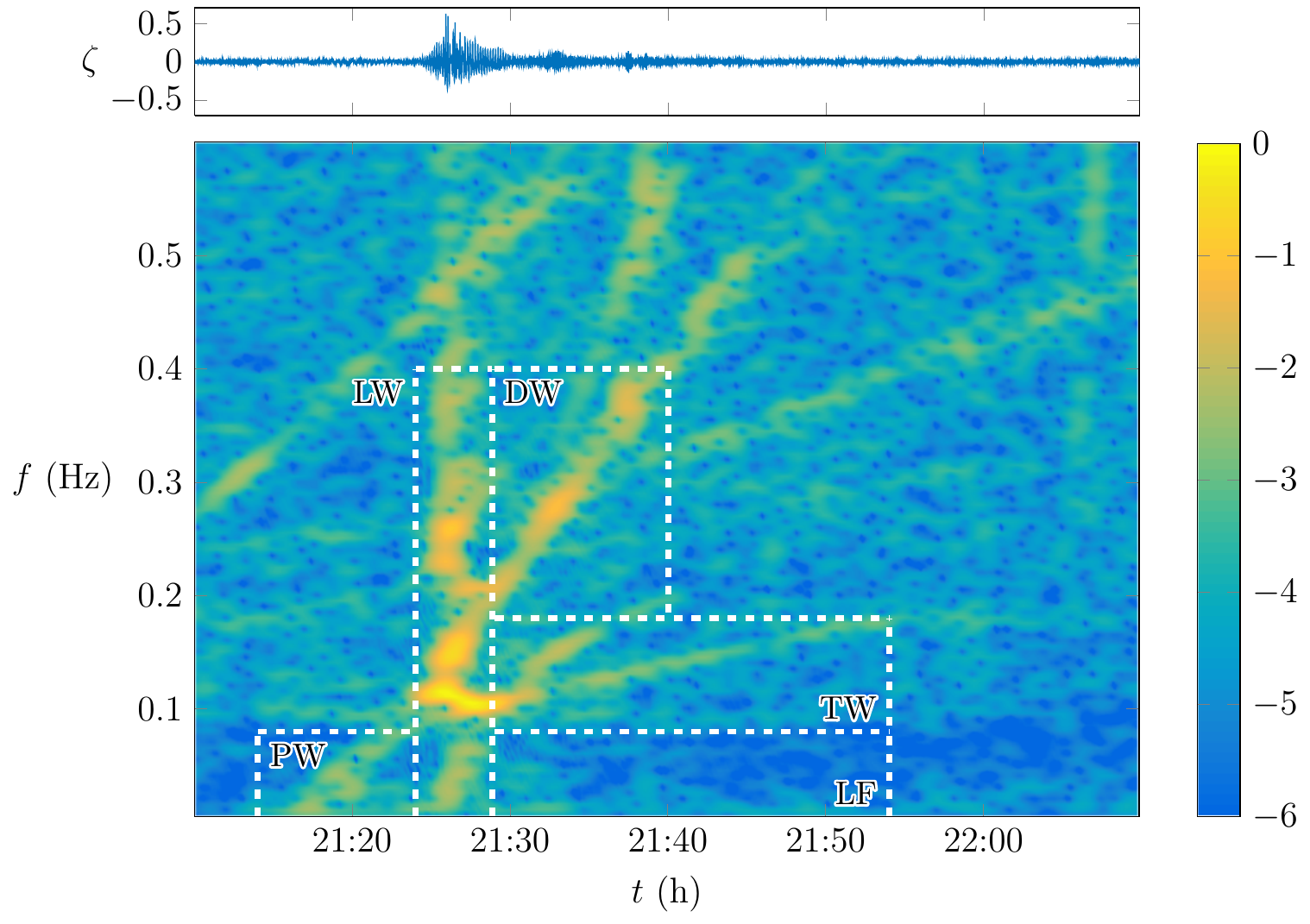}
\caption{A spectrogram of the signal generated by a high speed ferry in the Gulf of Finland.  This figure is identical to Figure 3(a) of \citet{didenkulova13}, except a) we have re-computed the spectrogram using a different colour scheme, { and b) we have used the colour intensity interval $(-6,0)$ whereas they use $(-6,-1)$}.  The five wake components as identified by \citet{torsvik15a} are the precursor {solitary} (PW), leading (LW), divergent (DW), transverse (TW) and low frequency (LF) waves. { The wave signal is presented above the spectrogram.}}
\label{fig:torsvikspec}
\end{figure}

By applying linear water wave theory, \citet{torsvik15a} showed that for small amplitude waves, the spectrogram of a steadily moving vessel has two linear components: a sliding-frequency mode (chirp) and a constant-frequency mode corresponding to divergent and transverse waves, respectively. The transverse and divergent components of the spectrogram were used to predict the ship's speed and the minimum distance from the echo sounder. However, by analysing high speed ferry data from the Gulf of Finland, \citet{torsvik15a} found and classified five wake components present in the spectrogram, the two linear components just mentioned plus three more; they referred to the additional three components as precursor {solitary}, leading and low frequency waves (Figure \ref{fig:torsvikspec}). \citet{torsvik15b} were able offer some evidence to the notion that two of the additional wake components (the so-called precursor and leading waves) were a result of wave shoaling and nonlinear effects caused by the ship waves approaching the shore.

\citet{didenkulova13} offered their own explanations for the features of the experimental spectrogram in Figure \ref{fig:torsvikspec}. They agree with \citet{torsvik15a} with respect to the cause of the precursor solitary wave and the leading wave being attributed to the nonlinear effects of the ship hull and wave shoaling. However, \citet{didenkulova13} attributed the two branches of colour intensity in the transverse wave component (in the boxed labelled TW in Figure~\ref{fig:torsvikspec}) to the transverse and divergent waves, stating that the transverse wave component represents the entire Kelvin wake signal. Finally, \citet{didenkulova13} did not provide an explanation for the features in the divergent and low frequency wave components.

The motivation for the present study is to use theoretical and computational methods to { identify linear and nonlinear features of idealised spectrograms and to} better understand the origin of the wake components in experimental spectrograms such as those presented by \citet{didenkulova13} and \citet{torsvik15a,torsvik15b}.

We begin by taking spectrograms of linear free-surface profiles to identify the two wake components that are present for linear flows past a pressure distribution (namely, the sliding-frequency mode and the constant-frequency mode).  A geometric argument is provided to show that these components are present for all linear ship wave patterns.  An advantage of using the idealised problem of flow past a pressure distribution (as opposed to flow past a single point pressure, as used by \citet{torsvik15a}), is that we are able to demonstrate how the high intensity signal in the spectrogram follows the constant-frequency mode (sliding-frequency mode) for slower (faster) ships.  We then modify the numerical method for computing nonlinear ship waves by \citet{pethiyagoda14a} to significantly increase the domain size of the computed solution. We use the numerical solutions to generate accurate spectrograms of nonlinear ship waves; these spectrograms are free of wind waves that are present in experimental measurements, allowing us to more easily observe the effects of steep nonlinear waves.  We identify features present in nonlinear spectrograms and use a weakly nonlinear theory to { derive analytical results that match well with the numerical simulations.  While it is very difficult to separate the direct influence of the high speed ferry in the spectrogram in Figure~\ref{fig:torsvikspec} from other effects or artifacts that may be unrelated to that particular vessel, we propose that nonlinear waves may be associated with} the leading wave component recorded by \citet{torsvik15a}.  Finally, we see the transverse wave component in Figure~\ref{fig:torsvikspec} is not horizontal, as predicted by linear theory.  We { provide a possible explanation for} this discrepancy by accounting for ship acceleration in the linear model.

\section{Problem setup}
In order to simulate a wake left behind a moving ship, we consider the idealised problem of calculating the free surface disturbance created by a steadily moving pressure distribution applied to the surface of an infinitely deep body of water.  We suppose the pressure distribution is of a Gaussian type with strength $P_0$ and characteristic length $L$, and then formulate the mathematical problem in the reference frame of this moving pressure.  We nondimensionalise the problem by scaling all velocities by the speed of the pressure distribution, $U$, and all lengths by $U^2/g$, where $g$ is acceleration due to gravity. The governing equations are then
\begin{align}
\nabla^2\phi&=0& \text{for }z&<\zeta(x,y),\label{eq:laplace}\\
\label{eq:bern}
\frac{1}{2}|\nabla\phi|^2+\zeta+\epsilon p&=\frac{1}{2}& \text{on }z&=\zeta(x,y),\\
\phi_x\zeta_x+\phi_y\zeta_y&=\phi_z& \text{on }z&=\zeta(x,y),\\
\phi&\rightarrow x& \text{as }x&\rightarrow -\infty,\label{eq:upstream}
\end{align}
where $\phi(x,y,z)$ is the velocity potential, $\zeta(x,y)$ is the free-surface height, $\epsilon=P_0/(\rho U^2)$ is the dimensionless pressure strength, $\rho$ is the fluid density and $\epsilon p(x,y)$ is the pressure distribution. For the present study we will use the pressure distribution
\begin{equation}
p(x,y)=\mathrm{e}^{-\pi^2F^4(x^2+y^2)},\label{eq:pressure}
\end{equation}
where $F=U/\sqrt{gL}$ is the Froude number.  In this formulation, $F$ is the parameter that measures the speed of the moving pressure, while the pressure strength $\epsilon$ provides a measure of nonlinearity in the problem (the regime $\epsilon\ll 1$ is approximately linear). { Note that in computing the Bernoulli constant on the right-hand side of (\ref{eq:bern}), we have assumed the surface height $\zeta\rightarrow 0$ far upstream as $x\rightarrow -\infty$}.

The use of (\ref{eq:pressure}) to represent a ship is obviously extremely simplistic.  Other simple models include a pair of pressure distributions, one positive and one negative to represent the bow and stern waves of a ship \citep{noblesse14}, a thin-ship approximation for when the beam of the ship is much less than the length \citep{michell98} or a flat-ship approximation for when the draft is much less than the length \citep{maruo67,tuck75}.  However, as a first step, we have found the use of (\ref{eq:pressure}) particularly insightful as the disturbance has a well defined centre point to aid our geometrical arguments. Additionally, Gaussian pressure distributions are still frequently used to approximate the ship when analysing properties of the wake \citep{darmon14,ellingsen14,pethiyagoda15,li16}.

\section{Spectrograms of small amplitude ship waves}

\subsection{Exact solution to linear problem}
For weak pressure distributions, $\epsilon\ll 1$, the problem (\ref{eq:laplace})-(\ref{eq:pressure}) can be linearised.  The linearised version has the exact solution \citep{wehausen60}
\begin{align}
\zeta(x,y) = &-\epsilon p(x,y)+\frac{\epsilon}{2\pi^2} \int\limits_{-\pi/2}^{\pi/2}\,\int\limits_{0}^{\infty}\frac{k^2\tilde{p}(k,\psi)\cos(k[|x|\cos\psi+y\sin\psi])}{k-k_0}\,\,\mathrm{d}k\,\,\mathrm{d}\psi\notag\\
&-\frac{\epsilon H(x)}{\pi} \int\limits_{-\pi/2}^{\pi/2}k_0^2\tilde{p}(k_0,\psi)\sin(k_0[x\cos\psi+y\sin\psi])\,\,\mathrm{d}\psi,\label{eq:exactLinearInfinite}
\end{align}
where $\tilde{p}(k,\psi)=\exp(-k^2/(4\pi^2F^4))/(\pi F^4)$ is the Fourier transform of the pressure distribution (\ref{eq:pressure}), $H(x)$ is the Heaviside function and the path of { integration over $k$} is taken below the pole $k=k_0$, where $k_0=\sec^2\psi$ { (thus this integral can be interpreted as a Cauchy Principal Value integral plus half of the residue at $k=k_0$)}.

\subsection{Linear spectrogram}

A spectrogram of a ship wake is generated by first taking a cross-section of the wave surface at a constant value of $y$ to create a wave signal, $s(t)$, where $t=x$ by changing the reference frame to move with the uniform flow (recall the dimensionless speed is unity). The spectrogram data is then given by the square magnitude of a short-time Fourier transform:
\begin{equation}
S(t,\omega)=\left|\int\limits_{-\infty}^{\infty}h(\tau-t)s(\tau)\mathrm{e}^{-i\omega\tau}\,\mathrm{d}\tau\right|^2,\label{eq:spec}
\end{equation}
where the window function, $h(t)$, is an even function with compact support. In this paper we will use the Blackman-Harris 92dB window function \citep{harris78}. The results are placed in a time-frequency heat map of angular frequency $\omega$ against scaled time $t/y$ and colour intensity on a log scale, $\log_{10}(S(t,\omega))$.

\begin{figure}
\begin{tabular}{ccc}
\subfloat[$F=0.3$]{\includegraphics[width=.45\linewidth]{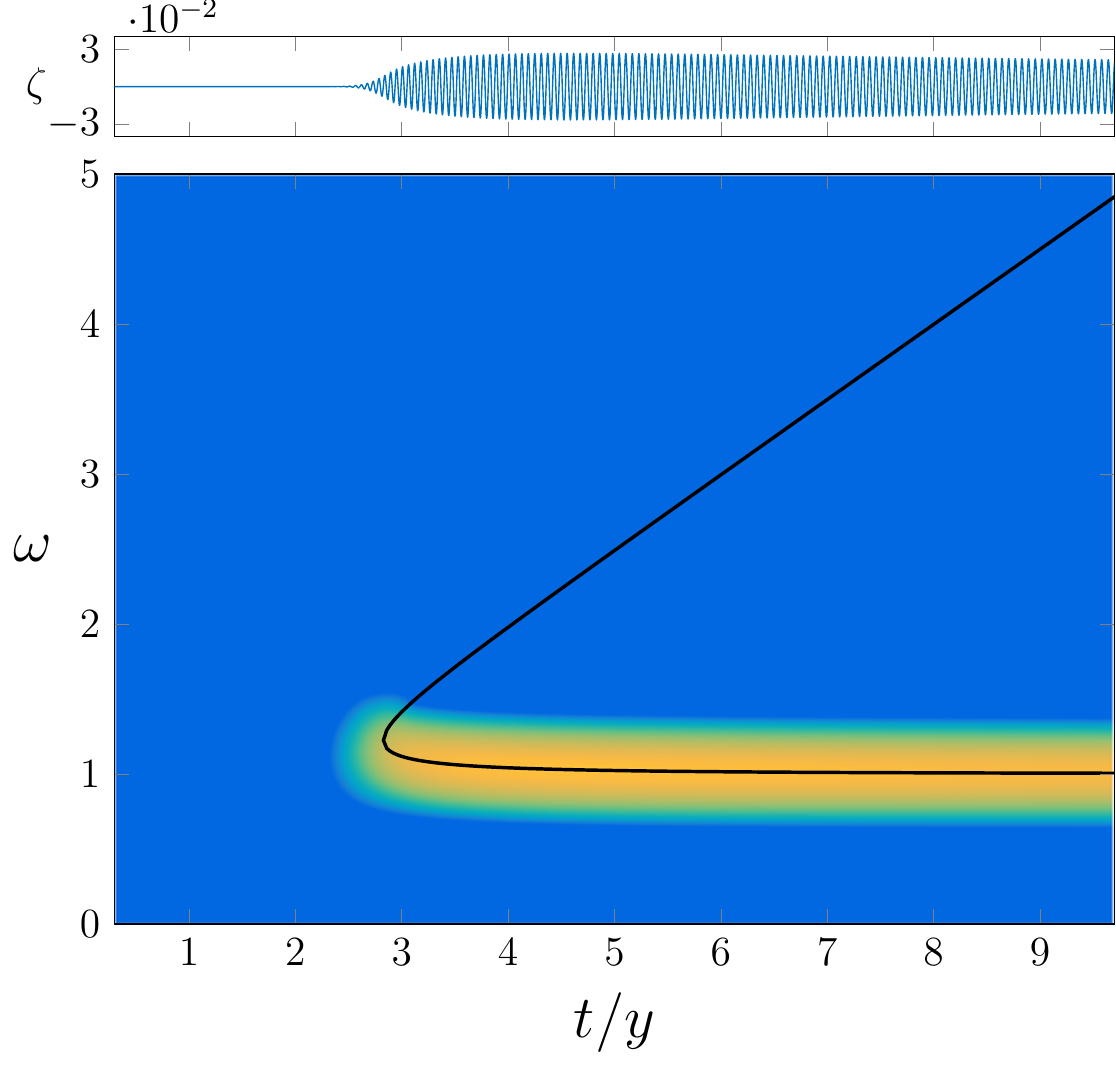}}&
\subfloat[$F=0.7$]{\includegraphics[width=.45\linewidth]{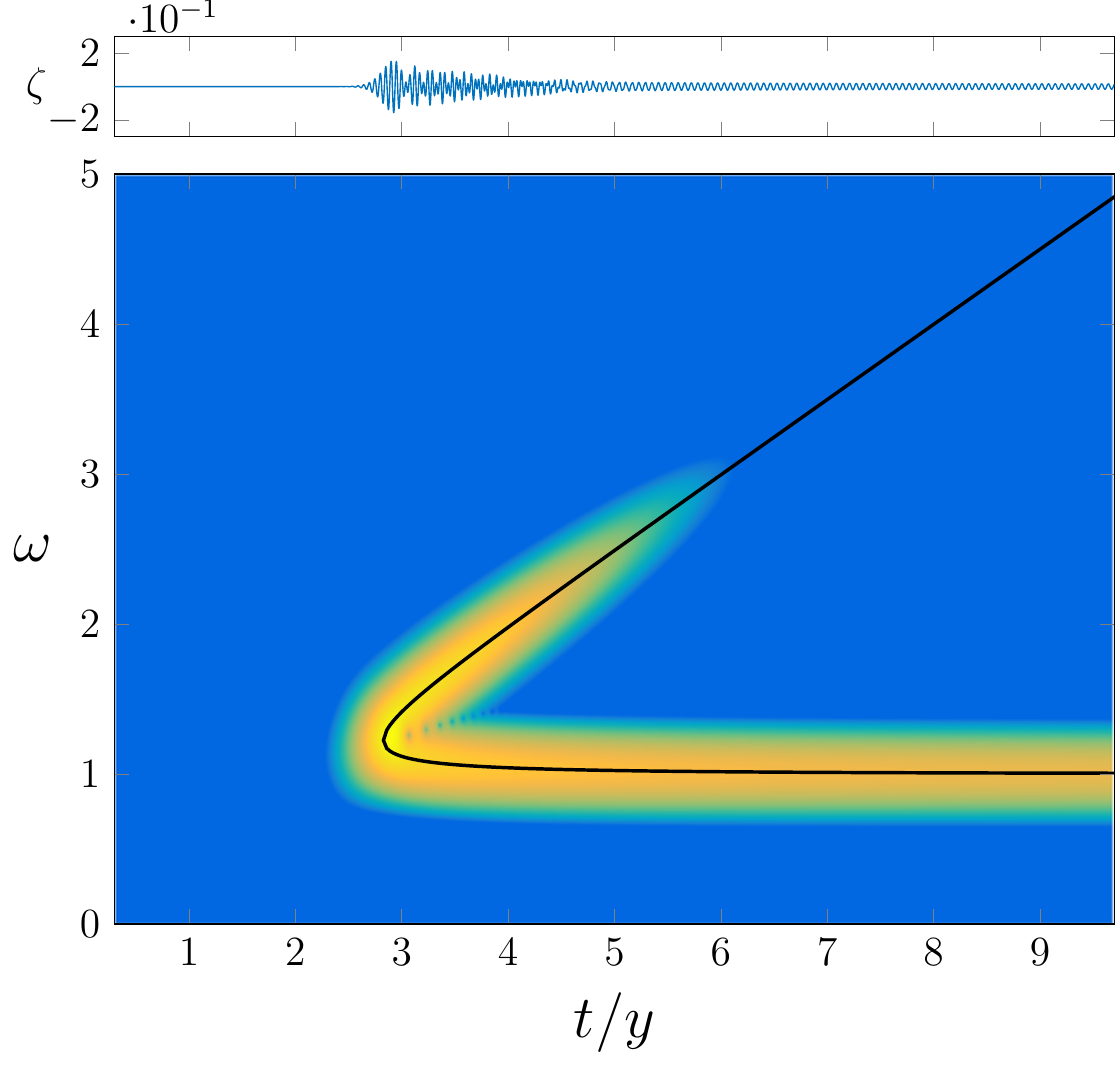}}&
\multirow{2}{*}[0.15\linewidth]{\includegraphics[width=.06\linewidth]{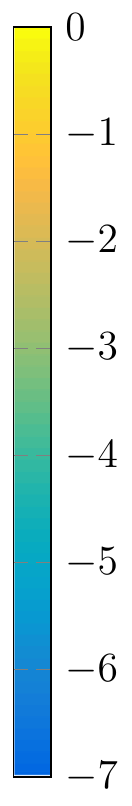}}\\
\subfloat[$F=1$]{\includegraphics[width=.45\linewidth]{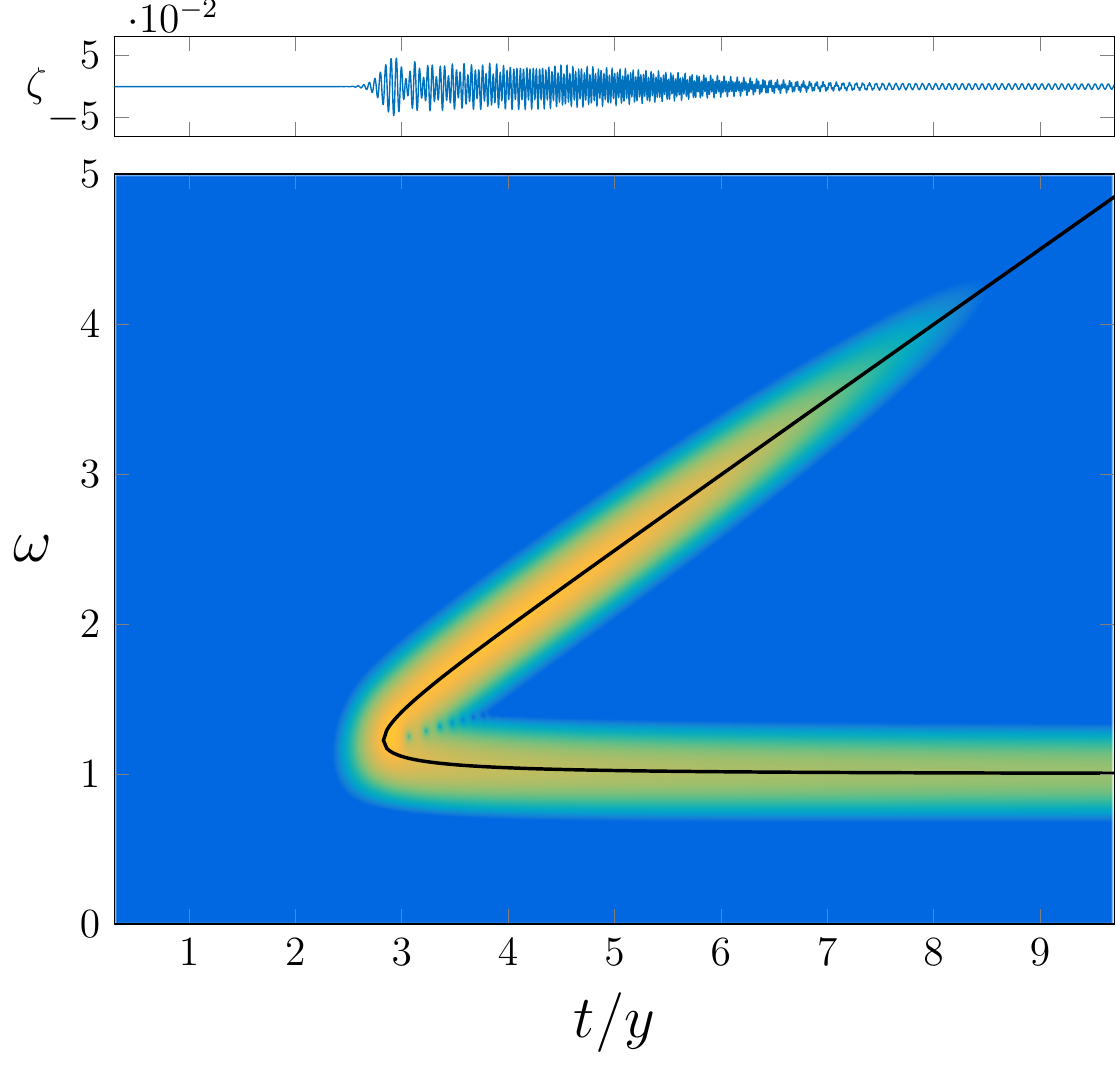}}&
\subfloat[$F=1.5$]{\includegraphics[width=.45\linewidth]{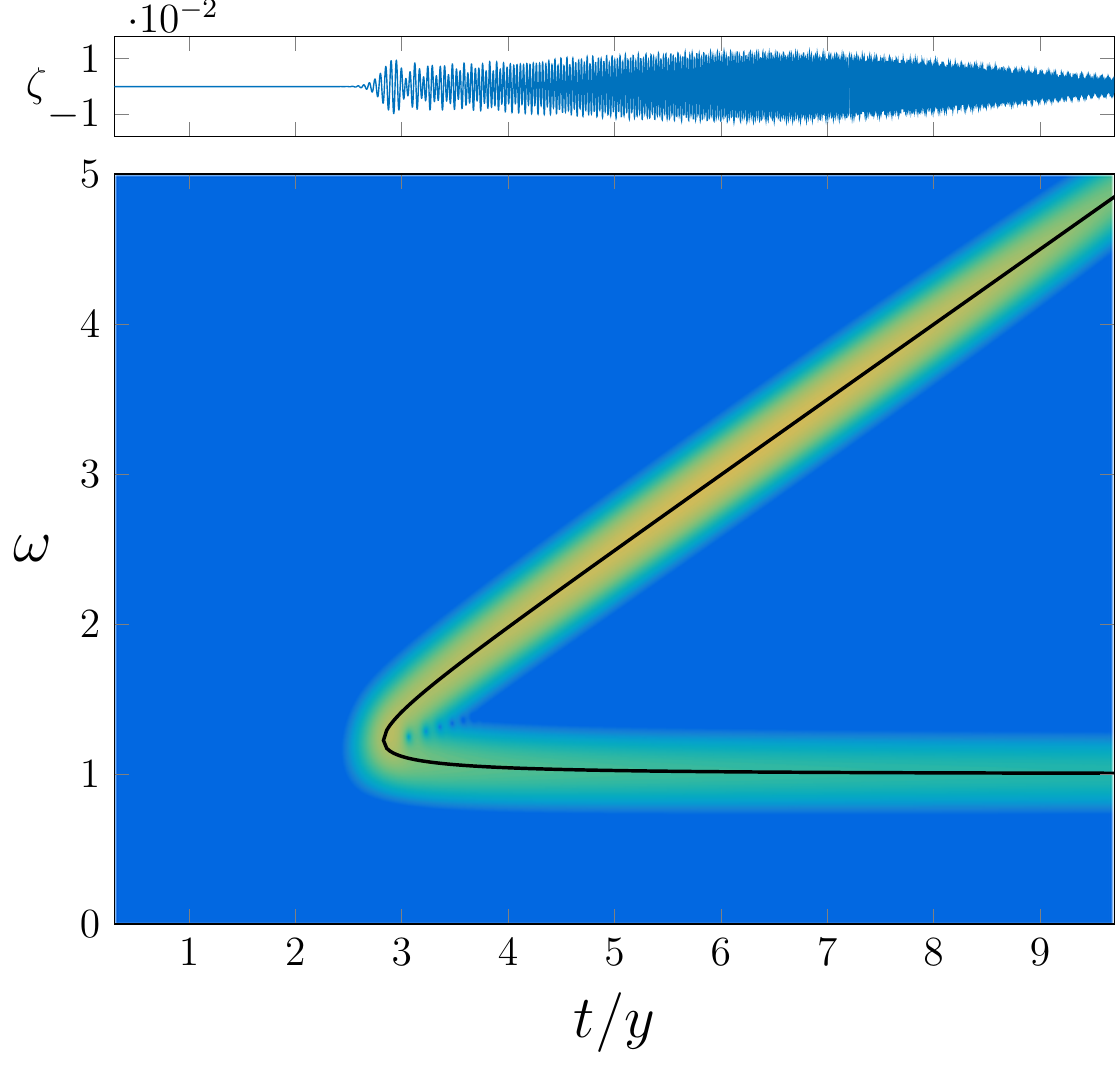}}&\\
\end{tabular}
\caption{Spectrograms of linear ship waves for the Froude numbers $F=0.3$, $0.7$, $1$ and $1.5$.  The solid curve is the dispersion curve given by equation (\ref{eq:specCurve}). The colour intensity is given by $\log_{10}(S(t,\omega))$, where $S(t,\omega)$ is given by (\ref{eq:spec}). { In each case, the wave signal $\zeta$, plotted against $t/y$, is shown above the related spectrogram.}}
\label{fig:linspec}
\end{figure}

Spectrograms computed from the linear solution (\ref{eq:exactLinearInfinite}) for the Froude numbers $F=0.3$, $0.7$, $1$ and $1.5$ are presented in Figure \ref{fig:linspec} together with a solid curve which we refer to as the linear dispersion curve.  More details are provided in the following subsection, but for now we note the linear dispersion curve has two branches, the lower branch corresponding to the transverse wave component of the spectrogram and the upper branch corresponding to the divergent wave component. { In this figure we have chosen to fix $y=100$, but note that for sufficiently large $y$ (say $y>25$), the spectrograms appear the same on this scale.  On the other hand, if $y$ is chosen to be too small (that is, the sample is too close to the pressure disturbance), there will be some unwanted blurring between the two branches in the spectrogram.}

There are two key features of the spectrograms in Figure \ref{fig:linspec}.  First, the high intensity portion of the spectrograms (the lighter coloured part) in parts (a)-(d) appears to be centred on the linear dispersion curve.  Thus, for this problem of linearised flow past a pressure distribution, the linear dispersion curve provides an excellent prediction for the dominant wave signals propagating past a representative point in space.  The second key feature is that the high intensity portion is confined to the lower branch of the dispersion curve for low Froude numbers (approximately $F<0.4$), while for larger Froude numbers it follows the fold and the upper branch.  This result is consistent with observations of ship wave patterns that suggest slowly moving ships give rise to wakes dominated by transverse waves, while fast vessels produce a wave train dominated by divergent waves.  We note that this second feature is not evident if we treat the idealised problem of flow past a point pressure (as in \citet{torsvik15a}), since the solution to the point-pressure problem does not depend on any parameter values apart from the pressure strength $\epsilon$ (there is no length scale and so no Froude number).

\begin{figure}
\centering
\includegraphics[width=.6\linewidth]{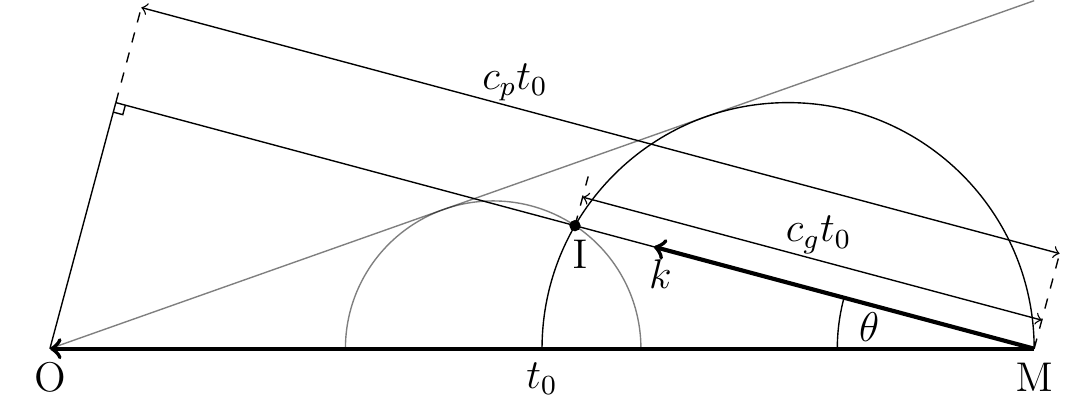}
\caption{This schematic of ship wave dispersion illustrates a ship moving from the point M to the point O with a nondimensional speed 1 over the time period $t_0$. The waves generated by the ship at the point M propagate at all angles $\theta$ from the sailing line and wavenumbers $k$ with phase velocity, $c_p$,  and group velocity, $c_g$. The large black semicircle is traced out using the group velocity for all $\theta$. The small grey semicircle follows the same rules but is generated at a different point. The grey line, tangent to the two semi circles, represents the geometric edge of the wake system. Finally, the point I is the location of the sensor.}
\label{fig:SpecDiagram}
\end{figure}

\subsection{Linear dispersion curve}

The linear dispersion curve in Figure \ref{fig:linspec} is independent of the speed of the disturbance (that is, independent of $F$) and can be determined via the following geometric arguments. First we consider a ship moving along the path MO a distance of $t_0$ as shown in Figure \ref{fig:SpecDiagram}. At the point M the ship will generate waves in all directions $\theta$ with phase velocity
\begin{equation}
c_p=\cos\theta\label{eq:phaseVel}
\end{equation}
and group velocity, $c_g$. To obtain the frequency at a given point $\mathrm{I}=(t,y)$, we must determine the angle $\theta$ and the associated wavenumber $k$. We require the following properties: an implicit function $k(\theta)$
\begin{equation}
k(\theta)^2\cos^2\theta-\Omega(k(\theta))^2=0,\label{eq:kFunc}
\end{equation}
where $\Omega(k)=\sqrt{k}$ is the dispersion function for waves on an infinite depth fluid; and a frequency relation given by projecting the wave vector $(-k\cos\theta,k\sin\theta)$  onto the direction the sensor is moving relative to the ship, $(1,0)$, and then taking the absolute value,
\begin{equation}
\omega = k(\theta)\cos\theta.\label{eq:omegaFunc}
\end{equation}
Finally, we use geometry to determine $\theta$, which can be found as a root of the equation
\begin{equation}
0=\tan^2\theta-\frac{t}{y}\alpha(k(\theta))\tan\theta+1-\alpha(k(\theta)),\label{eq:thetaZero}
\end{equation}
where $\alpha(k)=c_g/c_p=1/2$ is the ratio of group velocity to phase velocity in a fluid of infinite depth. Solving (\ref{eq:phaseVel})--(\ref{eq:thetaZero}) for $\omega$ gives
\begin{equation}
\omega_{1,2}=\frac{1}{2\sqrt{2}}\sqrt{\left(\frac{t}{y}\right)^2\pm \frac{t}{y}\sqrt{\left(\frac{t}{y}\right)^2-8}+4}.\label{eq:specCurve}
\end{equation}
The above argument is essentially equivalent to that provided by \citet{wyatt88}, although they do not provide the formula (\ref{eq:specCurve}).

The linear dispersion curve (\ref{eq:specCurve}) is the solid curve in Figure~\ref{fig:linspec}.  As mention above, it has two branches, the upper branch $\omega_1$ and lower branch $\omega_2$, and a fold where the two branches meet at the point ($t/y$, $\omega$)=($\sqrt{8}$,$\sqrt{3/2}$). The upper branch corresponds to the divergent waves; it approaches the line $\omega = t/2y$ for large $t/y$ and is represented by the smaller grey circle in Figure \ref{fig:SpecDiagram}.  The lower branch, represented by the large black circle in Figure \ref{fig:SpecDiagram}, corresponds to the transverse waves and approaches $\omega=1$ for large $t/y$. The fold represents the wedge boundary of the wave train in this geometric representation, providing the well-known Kelvin's ship wake angle, $\mathrm{arctan}\,(1/\sqrt{8})$ \citep{kelvin87}.

\section{Nonlinear ship wakes}\label{sec:nonlinear}

\subsection{Numerical scheme}\label{sec:numscheme}

There are no known exact solutions to the fully nonlinear problem (\ref{eq:laplace})--(\ref{eq:pressure}).  To compute numerical solutions we reformulate the equations using a boundary integral method, and construct a system of nonlinear equations $\textbf{F}(\textbf{u})=\textbf{0}$ using collocation \citep{forbes89,parau02,parau07}.  The key feature of this system is that the Jacobian is fully dense, which provides challenges in terms of accuracy and run-time. We have revisited this scheme recently with a preconditioned Jacobian-free Newton-Krylov (JFNK) method \citep{pethiyagoda14a}. The preconditioner used in the JFNK method was the block-banded linear preconditioner.  The solution was computed over a mesh of $N$ points in the $x$-direction and $M$ points in the $y$-direction, hereafter written as $N\times M$, with mesh spacing in the $x$ and $y$-directions given by $\Delta x$ and $\Delta y$, respectively.  Unfortunately, the largest computed domain provided by this method does not provide sufficient resolution in the frequency domain of the spectrogram to produce a reasonably accurate computed spectrogram. To overcome this limitation we have developed a sub-domain stitching method described here.

\begin{figure}
\centering
\includegraphics[width=.8\linewidth]{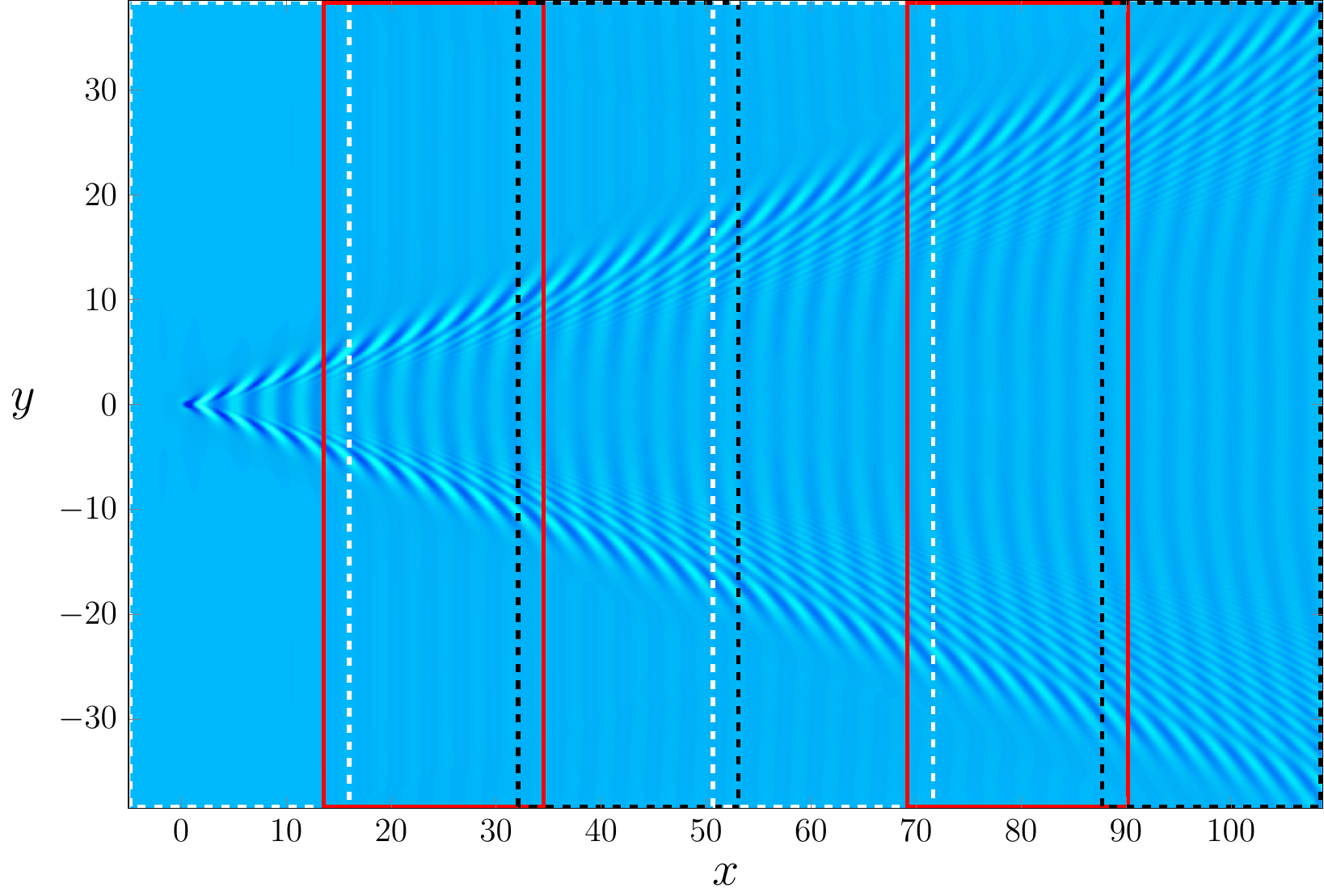}
\caption{Plan view of a nonlinear free-surface solution computed with $F=0.7$ and $\epsilon=0.01$ using the sub-domain stitching method. The solution is on a $1625\times 551$ mesh constructed from six $301\times 551$ panels indicated by the dashed {and solid} rectangles.}
\label{fig:StitchPlan}
\end{figure}

The sub-domain stitching method is an iterative procedure that computes sub-domains (hereafter referred to as panels) one at a time and connects them together, as shown in Figure \ref{fig:StitchPlan}. For clarity, subscripts are added to the parameters $N$, $M$, $\Delta x$ and $\Delta y$ denoting which panel they relate to. The first panel is computed on a $N_1\times M_1$ mesh that begins upstream of the pressure distribution as in the unaltered method, shown as the first white rectangle in Figure \ref{fig:StitchPlan}. { The first panel is then truncated from the downstream end onto a smaller $N_1^\prime\times M_1$ mesh, where $N_1^\prime<N_1$ is the number of points in the $x$-direction of the truncated domain. The truncation step is performed under the assumption that there exists domain truncation error near the downstream boundary.} The next panel is chosen such that $\Delta y_1=\Delta y_2$, $\Delta x_1=\Delta x_2$, $M_1\le M_2$ and is placed such that the upstream border of the second panel matches up with mesh points from the previous panel. { The second panel is placed such that the upstream points coincide with the downstream points of the truncated first panel} (first {solid} red rectangle in Figure \ref{fig:StitchPlan}). After the panel has been placed, the radiation conditions at the $M_1$ upstream points are changed from the algebraic decay of the original method \citep{pethiyagoda14a} to make sure there is continuity between panels { (ie. $\zeta_{N_1^\prime,j}^1=\zeta_{1,j}^2$, $\phi_{N_1^\prime,j}^1=\phi_{1,j}^2$, etc. for $j=1\dots M_1$ where $\zeta_{k,j}^i$ and $\phi_{k,j}^i$ are the surface height and velocity potential for the $k$th point in the $x$-direction and the $j$th point in the $y$-direction on the $i$th panel, respectively)}. The integral in the boundary integral equation can then be split into an integral over a known surface, the previously computed panel that does not overlap with the intended solution, and an integral over the unknown surface to be computed. This procedure is repeated with $\Delta y_1=\Delta y_i$, $\Delta x_1=\Delta x_i$ and $M_1\le M_2\le\cdots\le M_i$ until a satisfactory domain size is reached.  For example, Figure \ref{fig:StitchPlan} shows a free-surface profile corresponding to a solution computed with six panels. The preconditioner used in the JFNK method only needs to be updated for panel $i$ if $N_{i-2}=N_{i-1}=N_i$ and $M_{i-2}=M_{i-1}=M_i$ is not satisfied.

When computing highly nonlinear solutions it is prudent to use a bootstrapping method which takes the solution for a smaller value of $\epsilon$ as the initial guess to compute the solution for a larger value of $\epsilon$. This presents two methods of calculating solutions: compute the solution over all panels for a constant value of $\epsilon$ before increasing $\epsilon$ (panels first); or compute the solution for all desired $\epsilon$ values one panel at a time ($\epsilon$ first). Computing panels first has the advantage that it enables the full solution to be calculated for a given value of $\epsilon$ before continuing, as opposed to the $\epsilon$ first method which outputs the full solutions for all $\epsilon$ only at the end of the procedure. The $\epsilon$ first method will fail to return a solution over the full domain if one of the values of $\epsilon$ chosen does not have a solution (or takes unreasonably long to converge to the solution); however, computing $\epsilon$ first is faster as the preconditioner only needs to be updated when the panel is changed. Therefore, the choice of method depends on whether or not a solution is known to exist for all chosen values of $\epsilon$.

The main disadvantage of the stitching method described above is that spurious numerical waves are introduced at every boundary between panels. These numerical waves are roughly two-dimensional, small in amplitude, sinusoidal in nature, and more prominent for lower Froude numbers.  Due to these numerical waves,  a solution constructed via the stitching method will not be in complete numerical agreement with a single domain solution.  Spectrograms, however, are robust against such numerical waves. In fact, due to the sinusoidal nature of the numerical waves, they appear as a band of colour intensity at the constant value $\omega=1$ (that is, at the same frequency as the transverse waves along the centreline $y=0$) and so are readily identifiable and can be easily ignored.

\subsection{Second-order dispersion curves}

At this stage we have discussed the linear dispersion curve (\ref{eq:specCurve}) which does not account for nonlinear effects.  In order to include nonlinearity, we now derive an expression for further dispersion curves using a weakly nonlinear analysis. We follow the approach of \citet{hogben72}, who proposed a second-order solution to a ship wave problem provided by a finite number of $N$ monochromatic waves and their interaction terms,
\begin{equation}
\zeta(x,y)=\sum\limits_{n=1}^Na_n\cos R_n+\sum\limits_{r=1}^{N}\sum\limits_{s=1}^{r}\left\lbrace b_{rs+}\cos(R_r+R_s)+b_{rs-}\cos(R_r-R_s)\right\rbrace,\label{eq:secOrder}
\end{equation}
where the $a_n$ (for $n=1,\dots,N$) are the primary monochromatic wave amplitudes, $b_{rs+}$ and $b_{rs-}$ are non-zero amplitudes of the interacting waves and $R_n=k_n(x\cos\theta_n+y\sin\theta_n)$ defines the primary wave number $k_n$ and direction $\theta_n$.  By considering a solution which consists of only the two monochromatic waves ($N=2$) from linear theory, the amplitudes $a_{1,2}$ can easily be identified by comparing with the stationary phase approximation to the linear solution (\ref{eq:exactLinearInfinite}) which depends on the nondimensional parameters $\epsilon$ and $F$ { (the details of this calculation are not included here)}.  The wave number $k_n$ and direction $\theta_n$ are given by solving (\ref{eq:kFunc}) and (\ref{eq:thetaZero}) together to give
\begin{align}
\theta_{1,2}= & \tan^{-1}\left(\frac{(t/y)\pm\sqrt{(t/y)^2-8}}{4}\right),
\\
k_{1,2}= & \sec^2\theta_{1,2}=\frac{1}{8}\left[\left(\frac{t}{y}\right)^2\pm \frac{t}{y}\sqrt{\left(\frac{t}{y}\right)^2-8}+4\right].
\end{align}
The amplitudes of the interacting waves $b_{rs\pm}$ are proportional to $a_r a_s$.  For further details, see \citet{hogben72}.

To determine the location of the { additional} dispersion curves { that arise from the second-order solution (\ref{eq:secOrder})}, the actual wave amplitudes $a_n$ and $b_{rs\pm}$ are not required (provided they are non-zero).  { Instead,} by examining the { phases} of the interacting waves { (that is, $R_r+R_s$ and $R_r-R_s$)}, we can easily calculate the { additional} dispersion curves by either doubling the linear frequencies, adding them together, or taking the difference between them:
\begin{align}
\omega_{3,4} &= 2\omega_{1,2} = \frac{1}{\sqrt{2}}\sqrt{\left(\frac{t}{y}\right)^2\pm \frac{t}{y}\sqrt{\left(\frac{t}{y}\right)^2-8}+4},\label{eq:specCurve2}\\
\omega_{5,6} &= \omega_1 \pm \omega_2 = \frac{1}{2}\sqrt{\left(\frac{t}{y}\right)^2\pm 4\sqrt{\left(\frac{t}{y}\right)^2-1}+4}.\label{eq:specCurve3}
\end{align}
{ We refer to the curves described by  (\ref{eq:specCurve2})-(\ref{eq:specCurve3}) as being the second-order dispersion curves as they come from analysing the second-order solution (\ref{eq:secOrder}).}

\begin{figure}
\centering
\includegraphics[width=.5\linewidth]{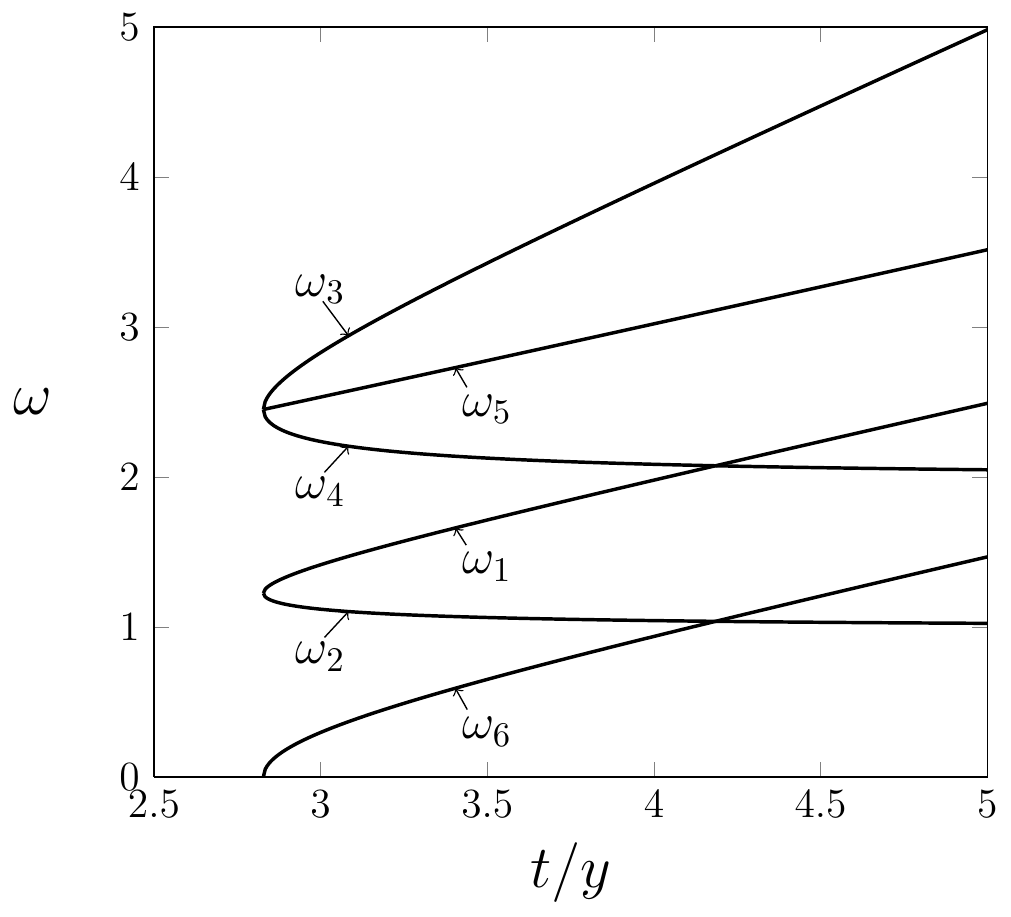}
\caption{The linear and second order dispersion curves. The linear dispersion curve (with branches $\omega_{1,2}$) is given by (\ref{eq:specCurve}). The second order dispersion curves (with branches $\omega_{3,4}$ and $\omega_{5,6}$) are given by (\ref{eq:specCurve2}) and (\ref{eq:specCurve3}), respectively.}
\label{fig:SecOrderDisp}
\end{figure}

The linear and second-order dispersion curves are shown in Figure \ref{fig:SecOrderDisp}.  We emphasise that, like their linear counterparts, these second-order dispersion curves do not depend on $\epsilon$ or $F$.  Instead, they indicate which parts of the time-frequency domain could be highlighted in a spectrogram for a nonlinear wave pattern.  Whether one or more of the branches are actually highlighted depends on the strength of the nonlinearity $\epsilon$ and the speed of the ship $F$, as we see in the following subsection.

\subsection{Spectrograms computed using our nonlinear simulations}

\begin{figure}
\begin{tabular}{ccc}
\subfloat[$\epsilon=0.15$]{\includegraphics[width=.45\linewidth]{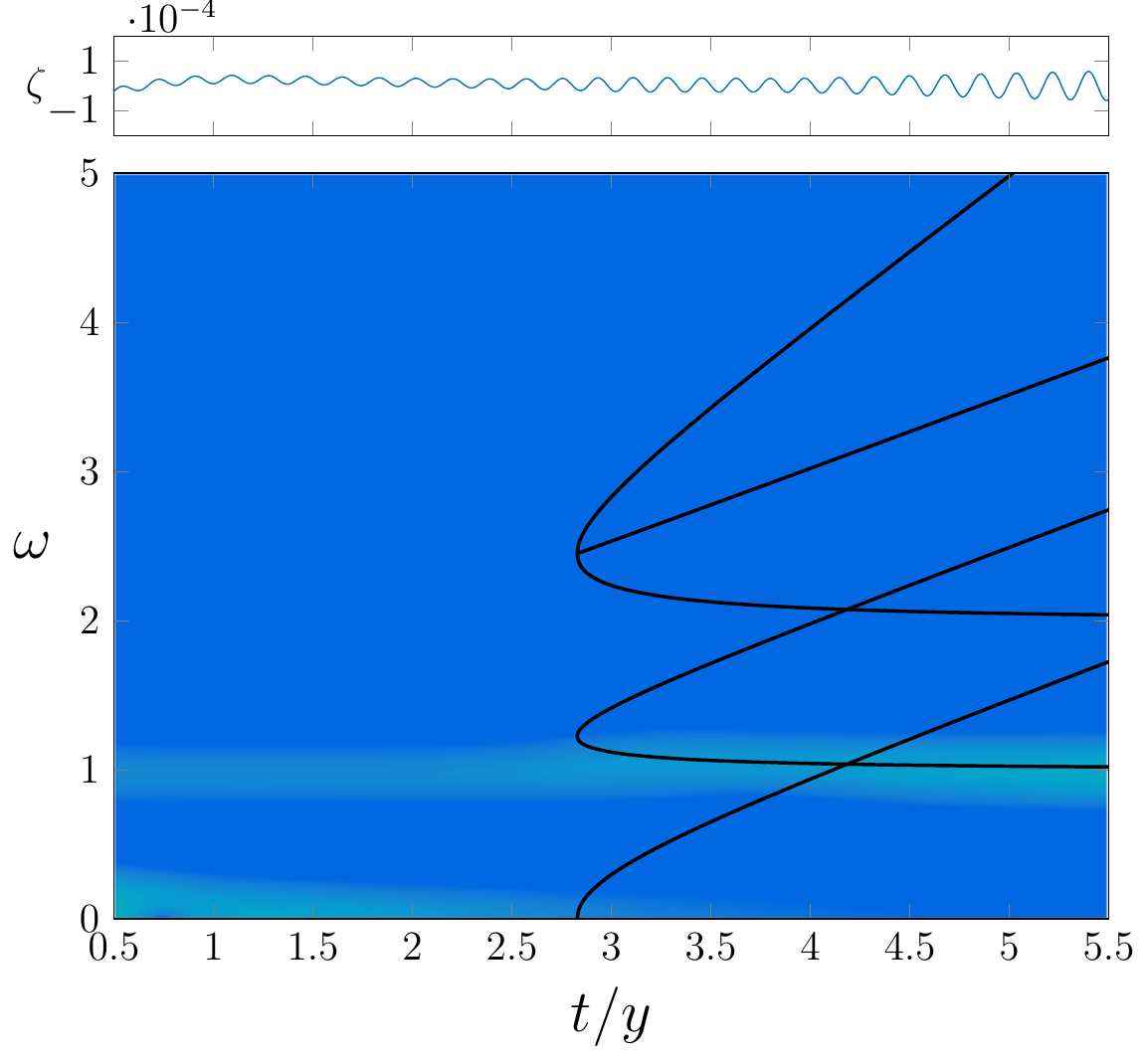}}&
\subfloat[$\epsilon=0.75$]{\includegraphics[width=.45\linewidth]{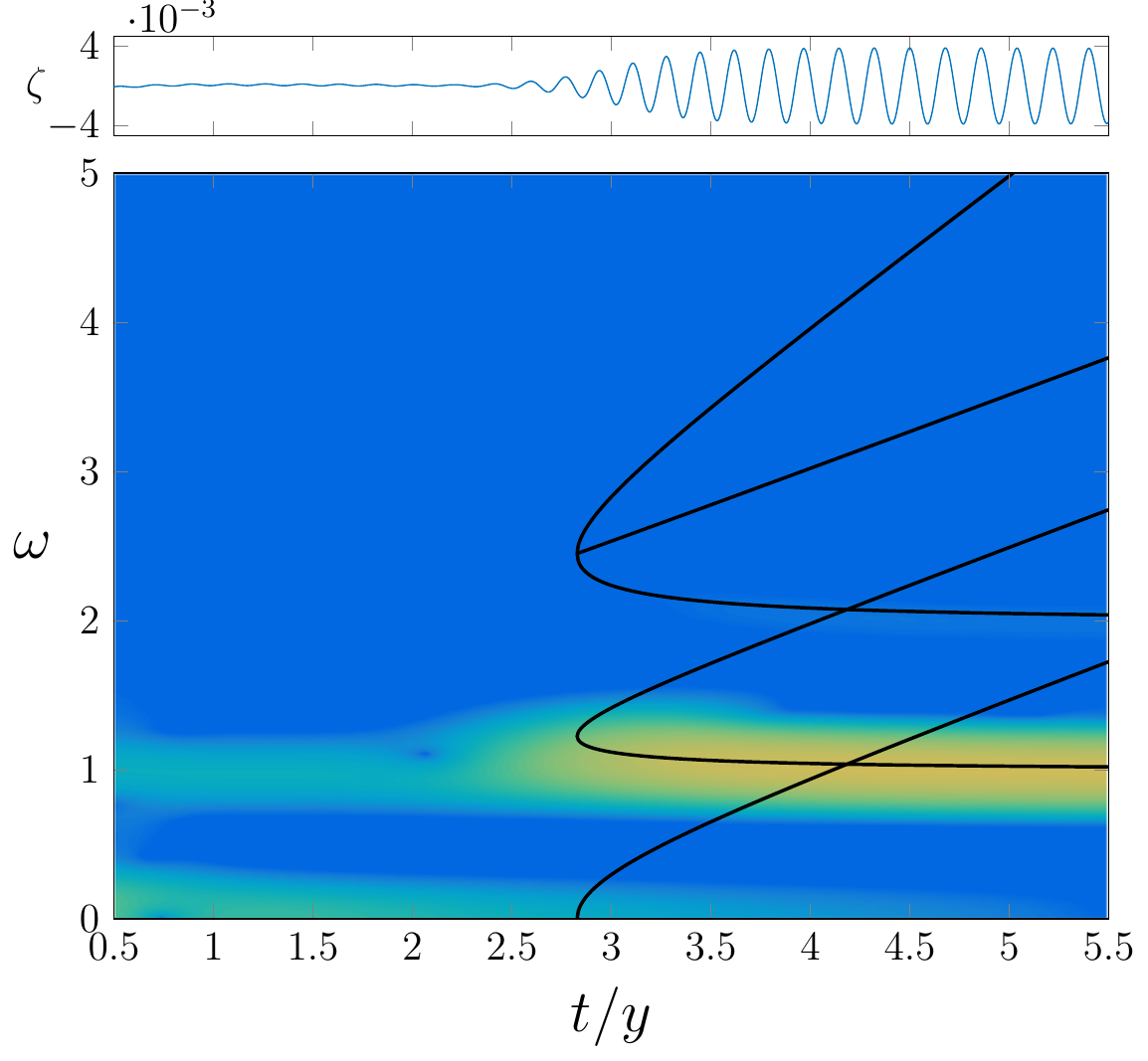}}&
\multirow{2}{*}[0.15\linewidth]{\includegraphics[width=.06\linewidth]{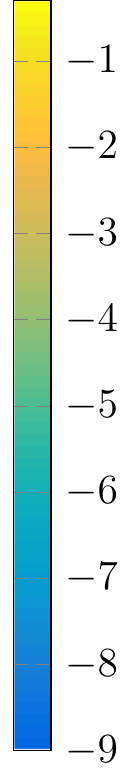}}\\
\subfloat[$\epsilon=1.65$]{\includegraphics[width=.45\linewidth]{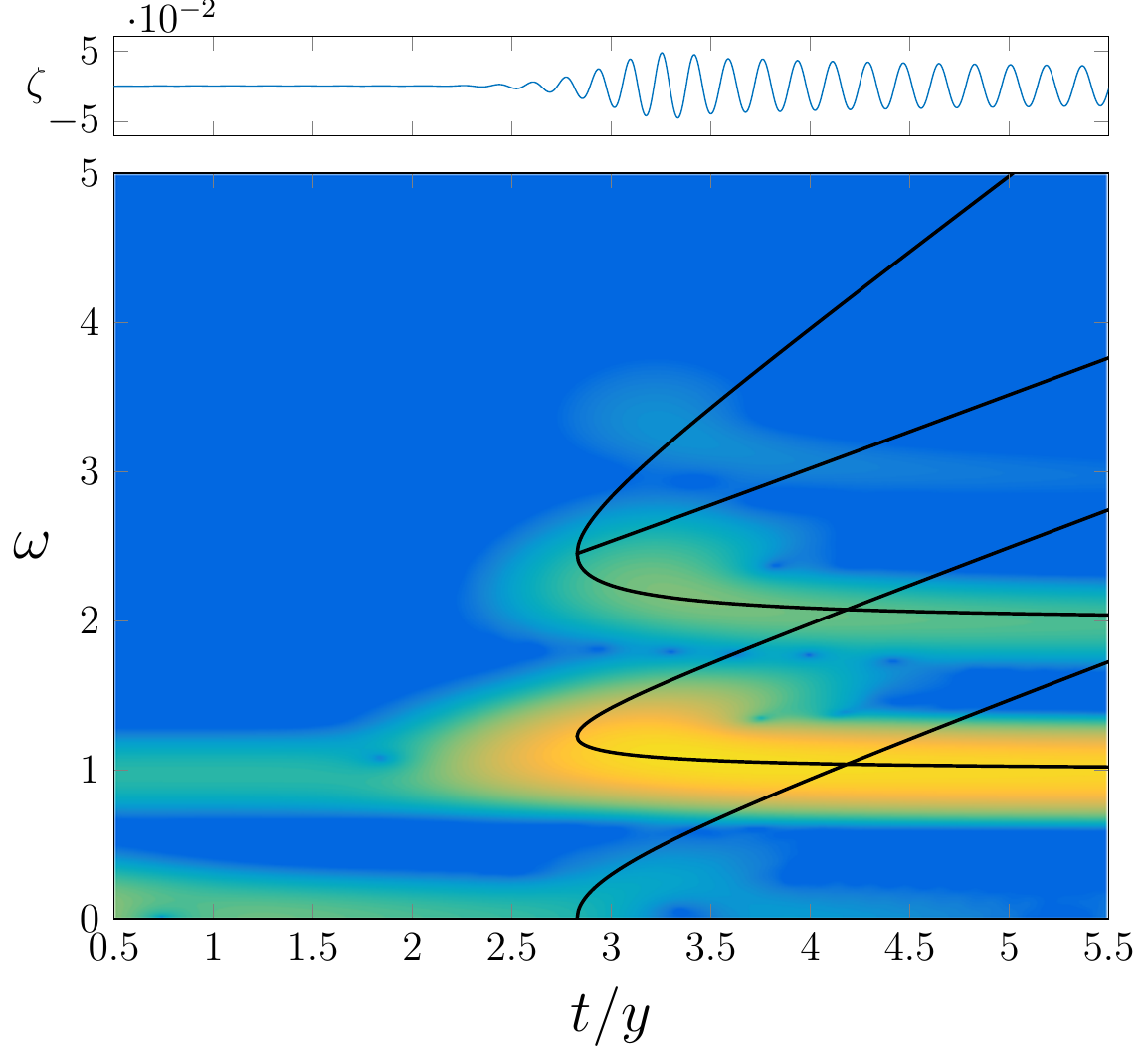}}&
\subfloat[$\epsilon=2.25$]{\includegraphics[width=.45\linewidth]{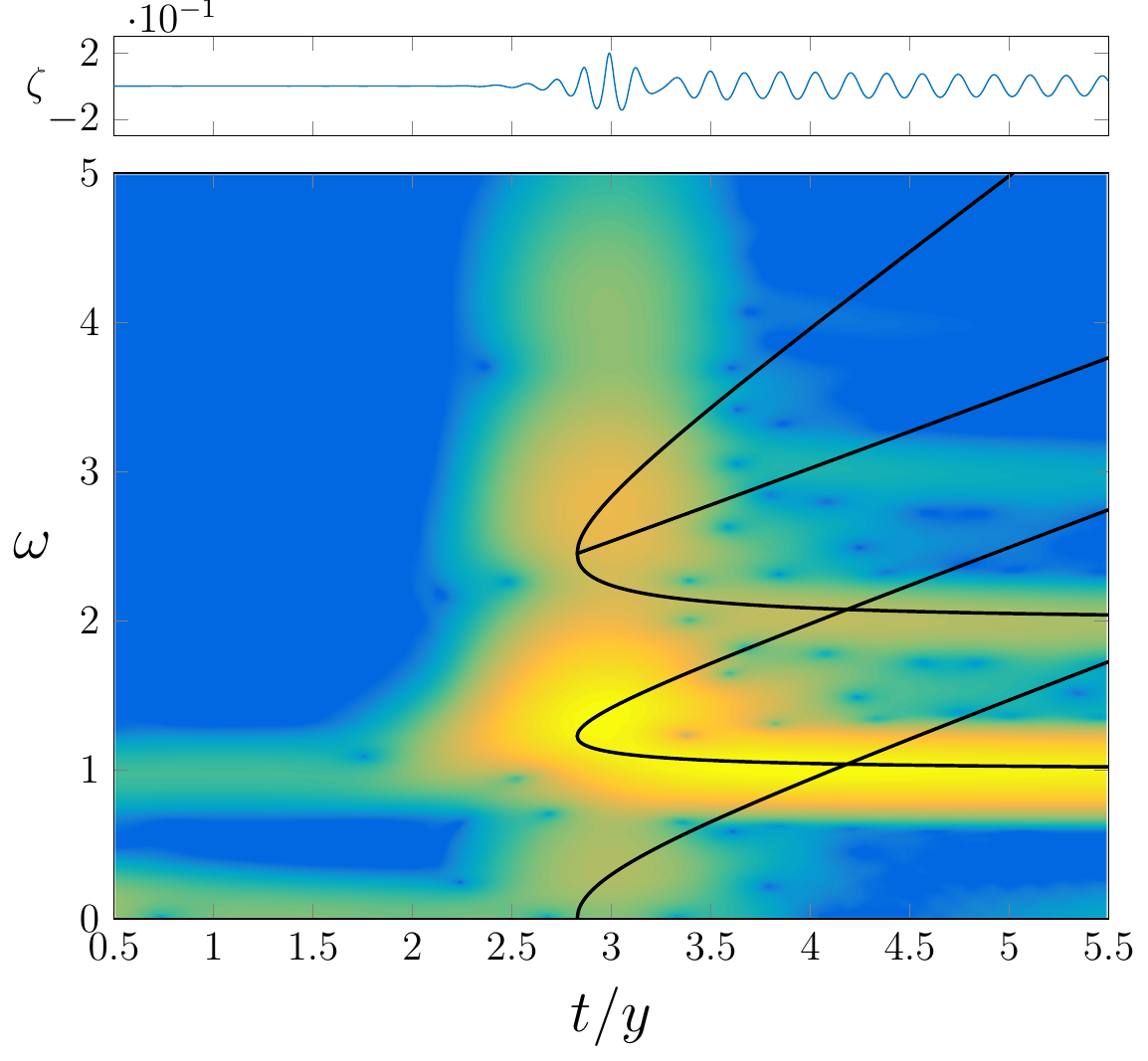}}&\\
\end{tabular}
\caption{Spectrograms of nonlinear ship waves for nondimensional pressure strength number $\epsilon=0.15$, $0.75$, $1.65$, $2.25$ and Froude number $F=0.2$. The solid curves are the linear and second-order dispersion curves given by equations (\ref{eq:specCurve}), (\ref{eq:specCurve2}) and (\ref{eq:specCurve3}). The colour intensity is given by $\log_{10}(S(t,\omega))$ where $S(t,\omega)$ is given by (\ref{eq:spec}). { In each case, the wave signal $\zeta$, plotted against $t/y$, is shown above the related spectrogram.}}
\label{fig:nonlinspecFr02}
\end{figure}

\begin{figure}
\begin{tabular}{ccc}
\subfloat[$\epsilon=0.01$]{\includegraphics[width=.45\linewidth]{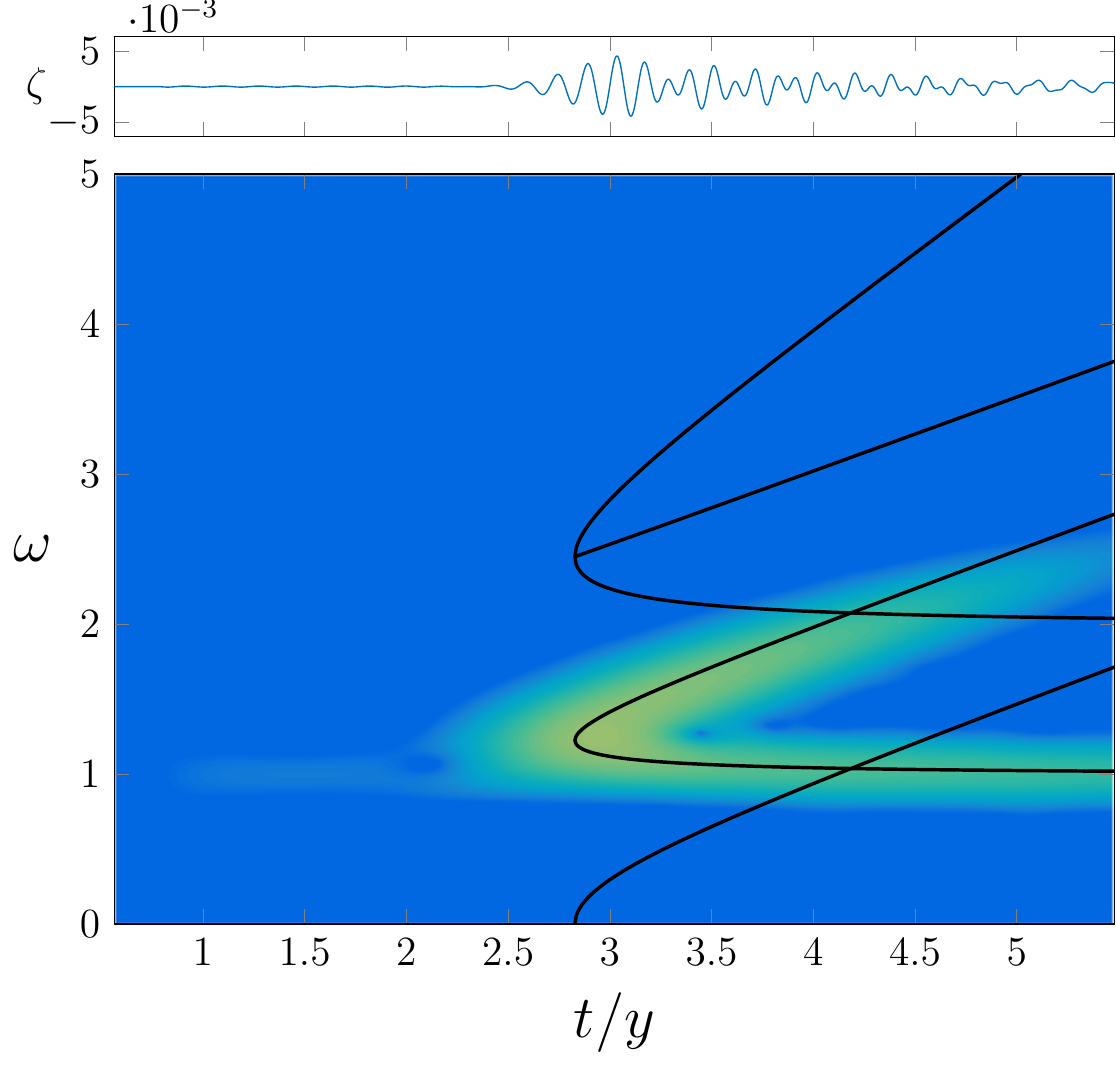}}&
\subfloat[$\epsilon=0.05$]{\includegraphics[width=.45\linewidth]{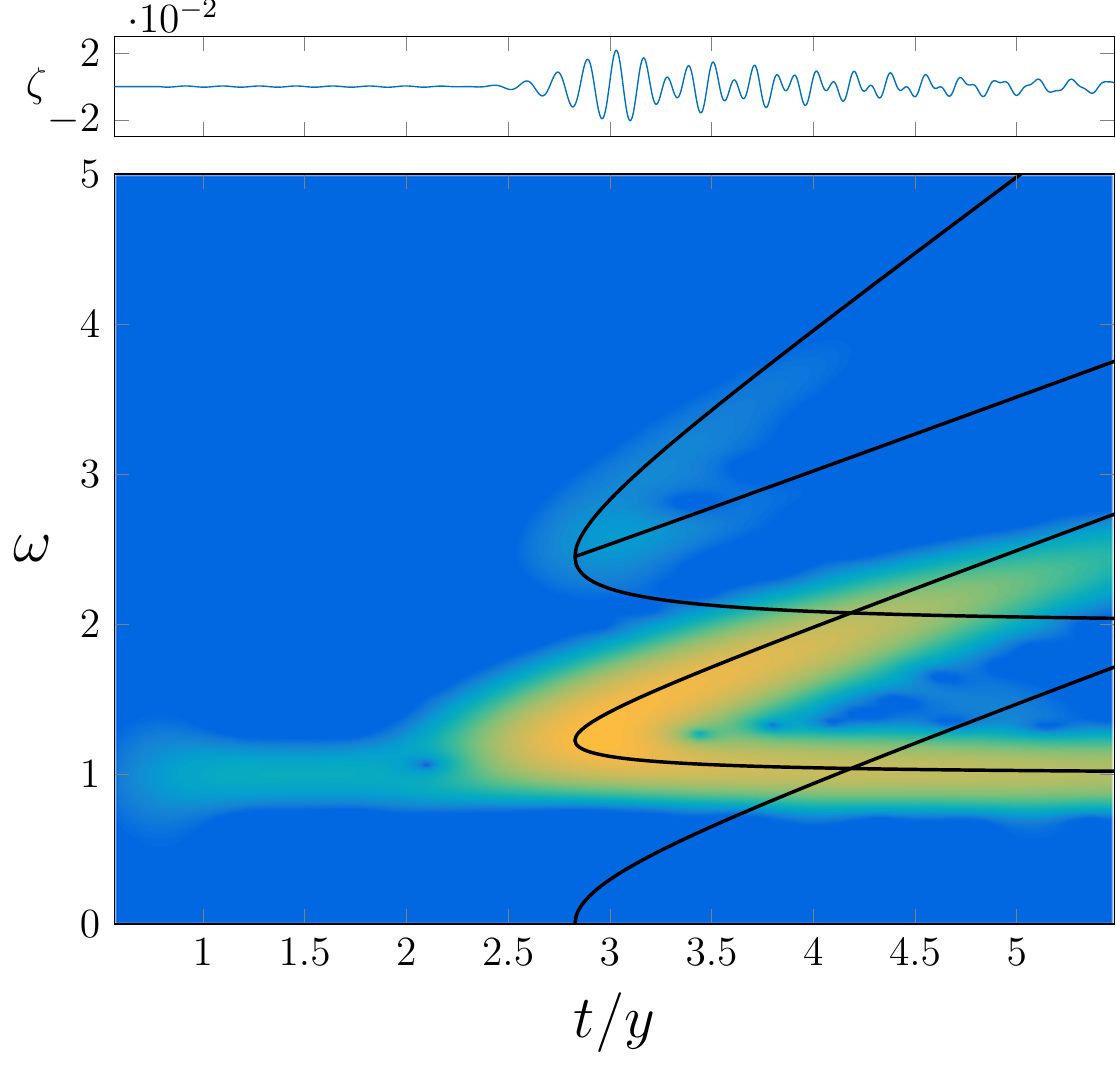}}&
\multirow{2}{*}[0.15\linewidth]{\includegraphics[width=.06\linewidth]{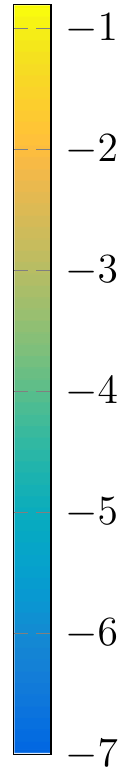}}\\
\subfloat[$\epsilon=0.11$]{\includegraphics[width=.45\linewidth]{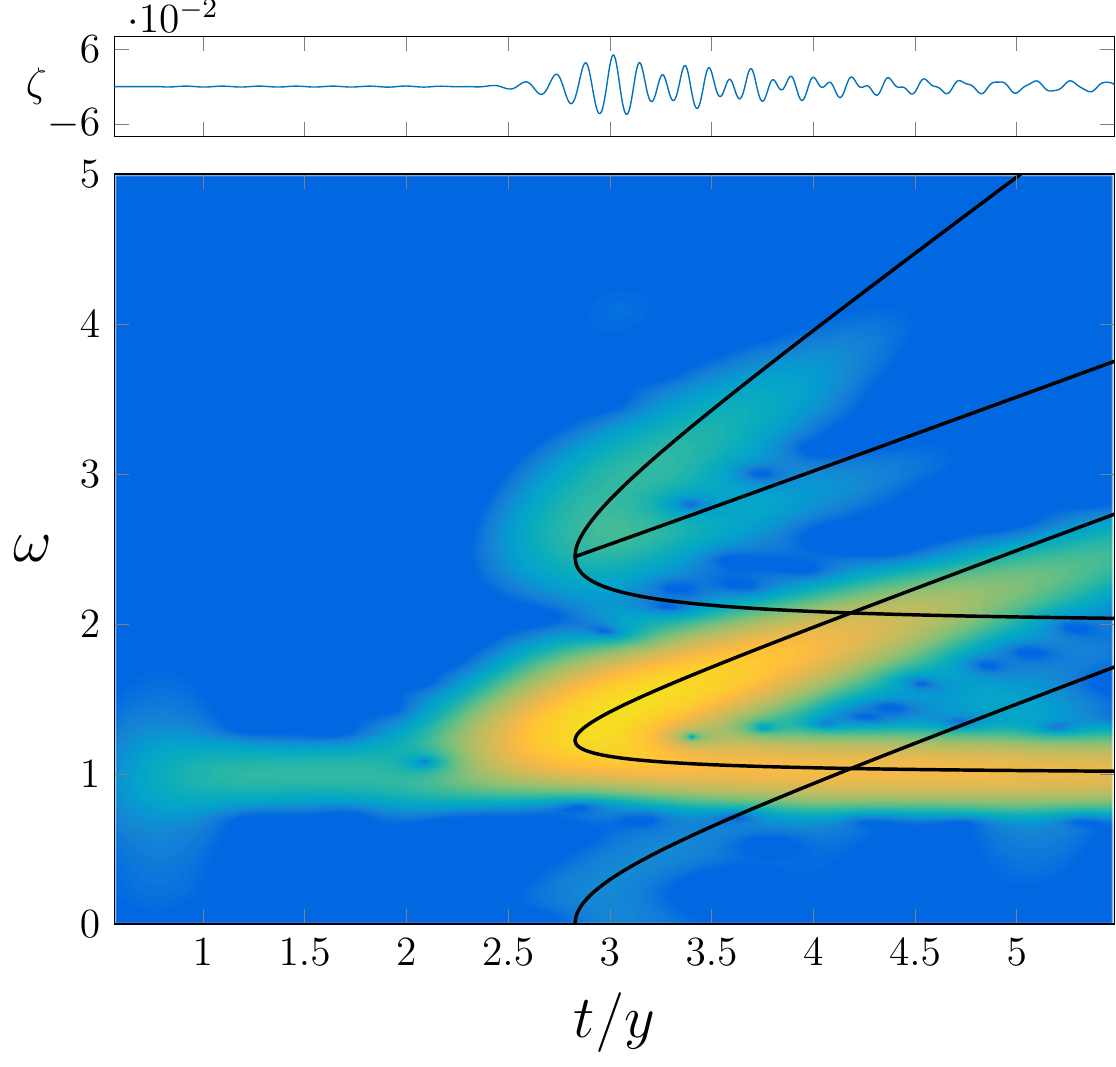}}&
\subfloat[$\epsilon=0.15$]{\includegraphics[width=.45\linewidth]{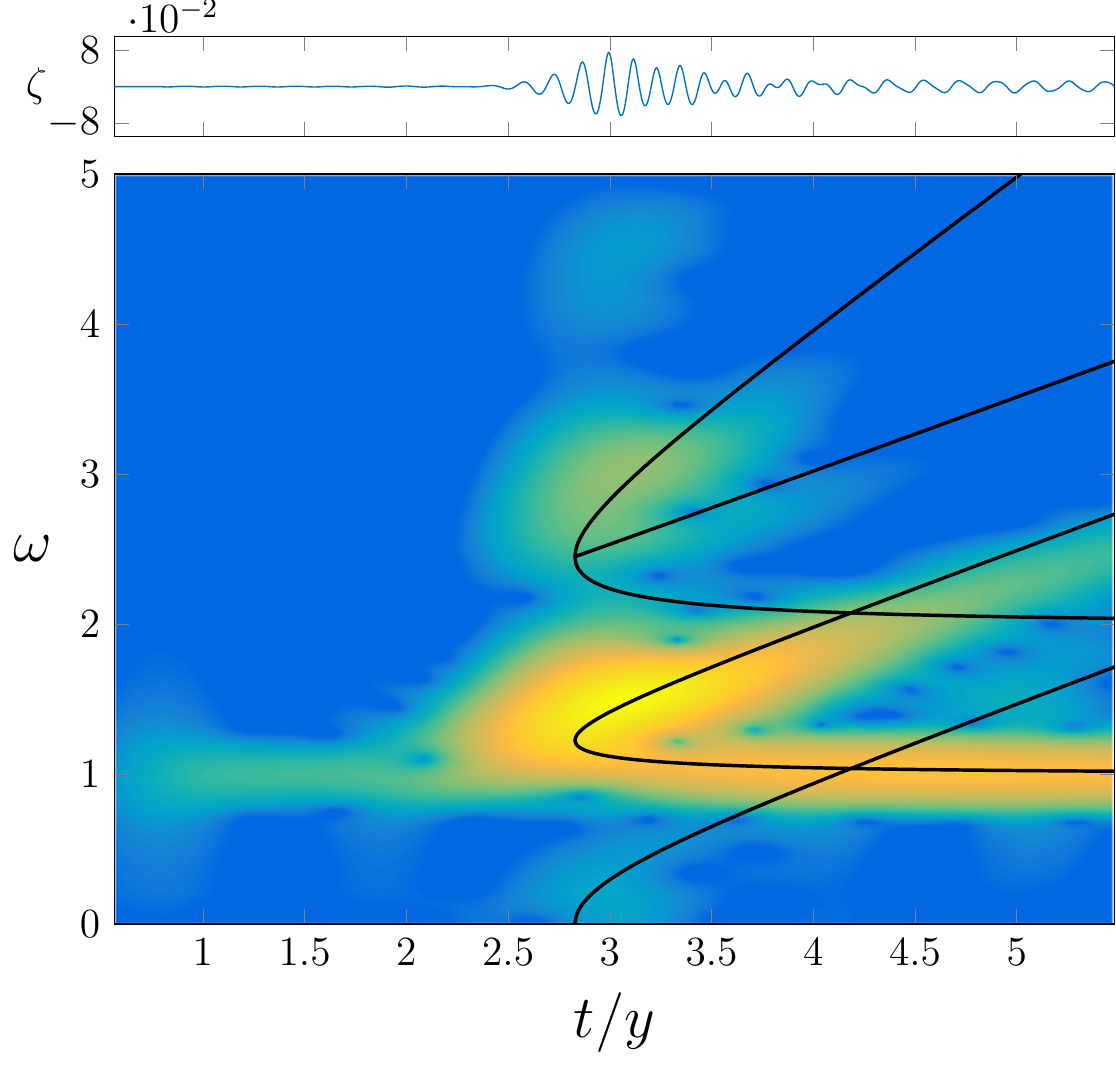}}&\\
\end{tabular}
\caption{Spectrograms of nonlinear ship waves for nondimensional pressure strength number $\epsilon=0.01$, $0.05$, $0.11$, $0.15$ and Froude number $F=0.7$. The solid curves and colour intensity are the same as in Figure~\ref{fig:nonlinspecFr02}.}
\label{fig:nonlinspecFr07}
\end{figure}

\begin{figure}
\begin{tabular}{ccc}
\subfloat[$\epsilon=0.01$]{\includegraphics[width=.45\linewidth]{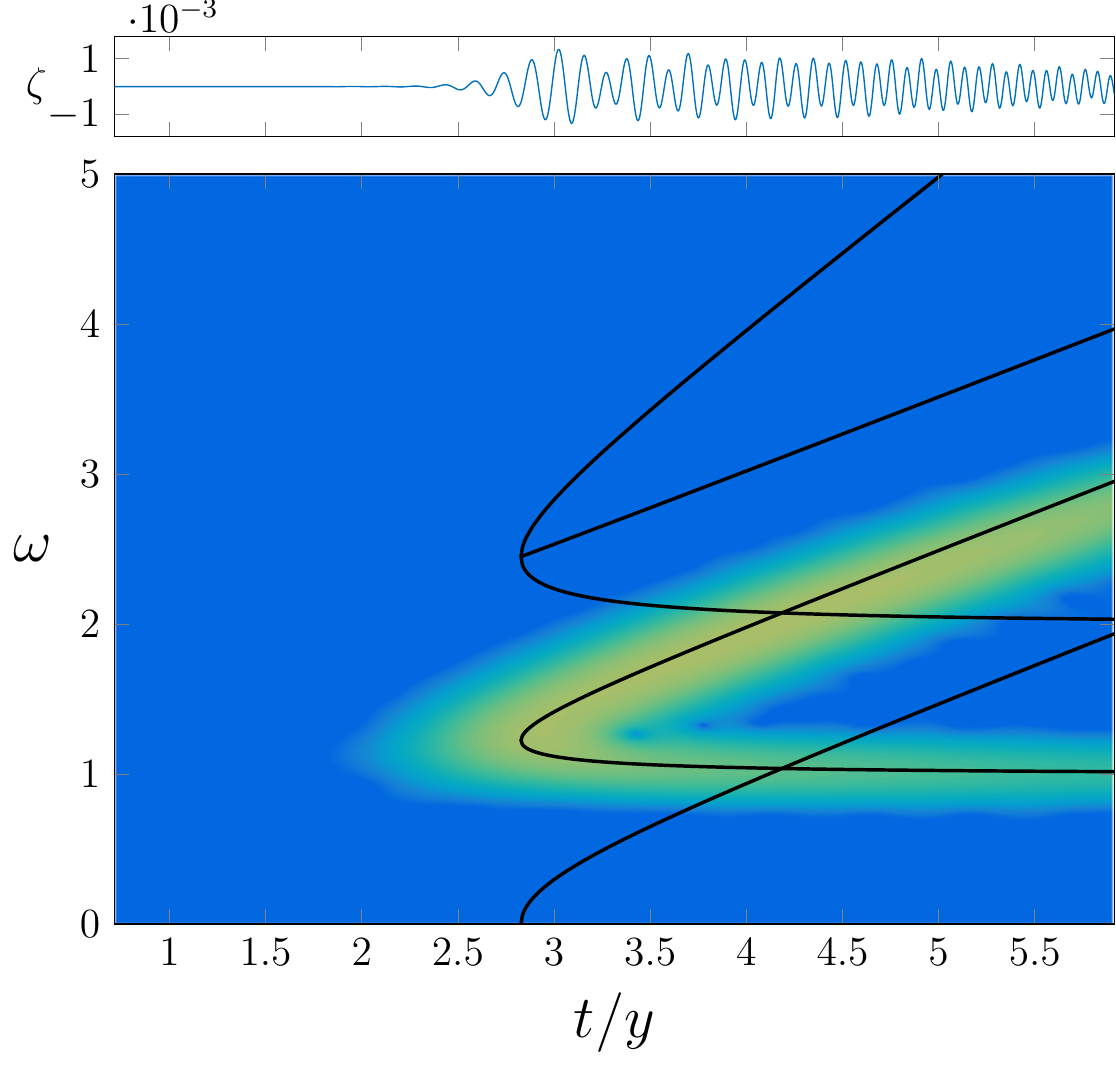}}&
\subfloat[$\epsilon=0.05$]{\includegraphics[width=.45\linewidth]{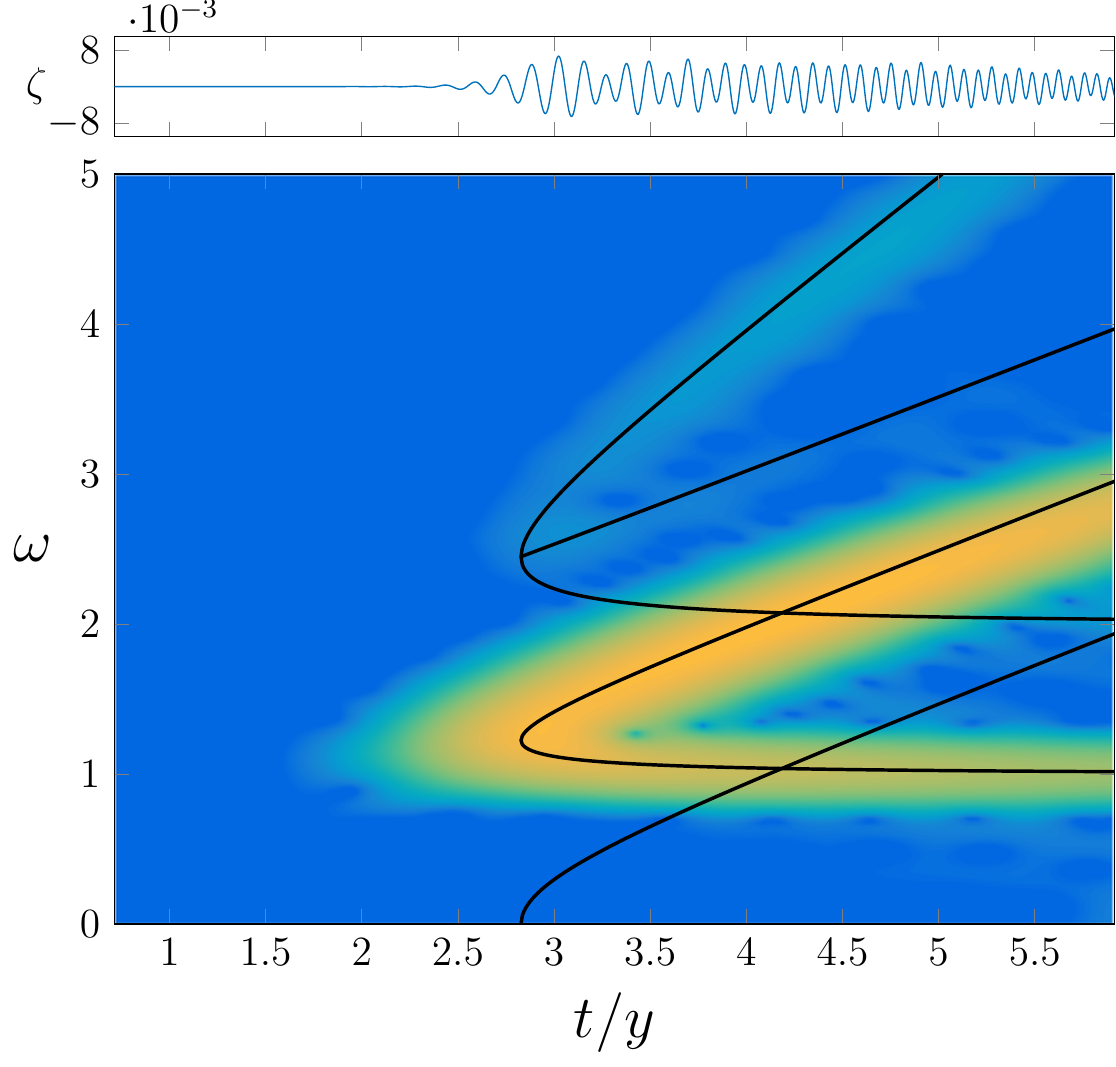}}&
\multirow{2}{*}[0.15\linewidth]{\includegraphics[width=.06\linewidth]{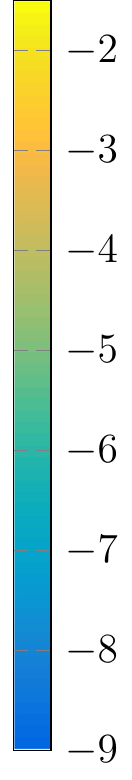}}\\
\subfloat[$\epsilon=0.11$]{\includegraphics[width=.45\linewidth]{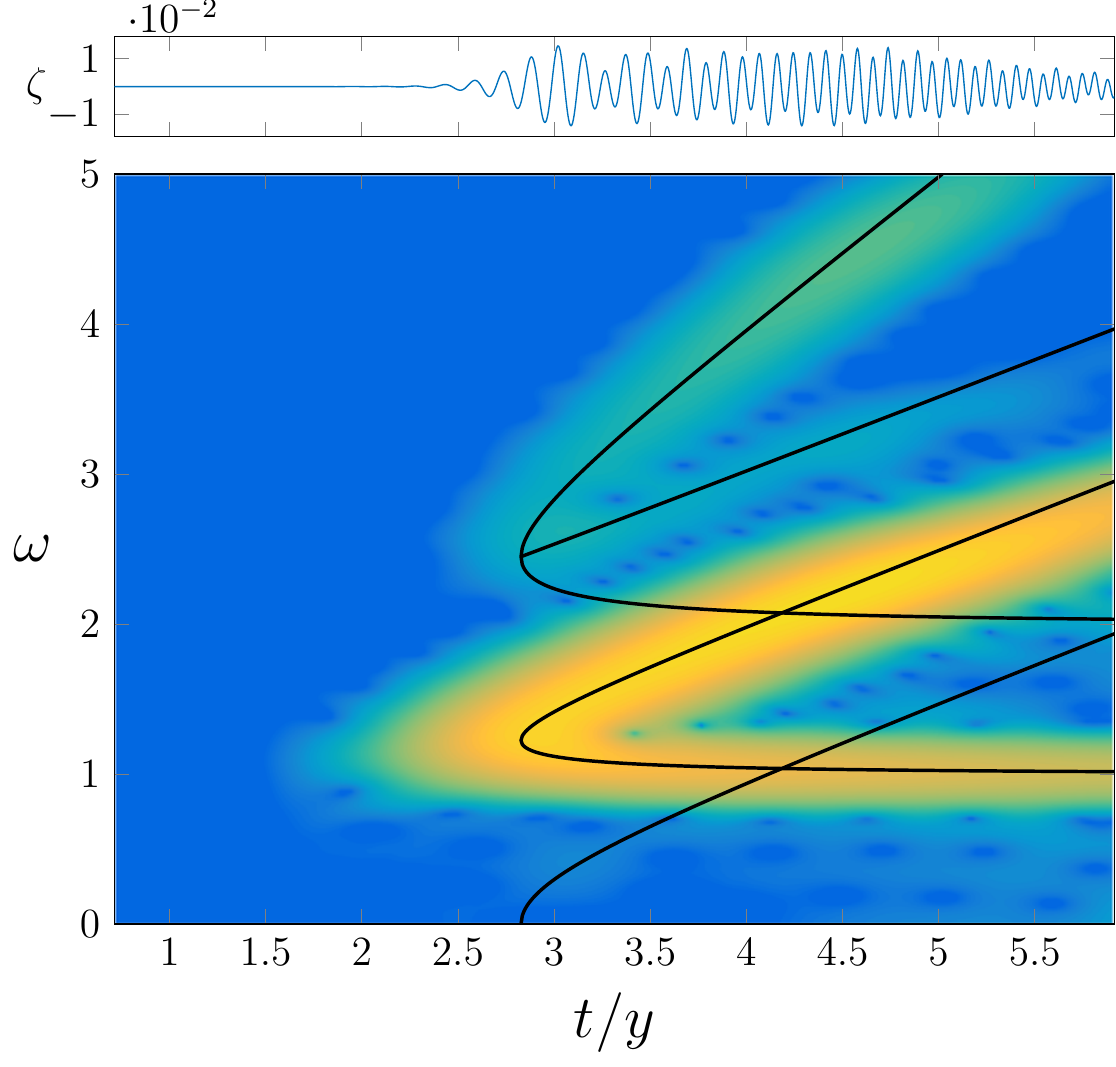}}&
\subfloat[$\epsilon=0.15$]{\includegraphics[width=.45\linewidth]{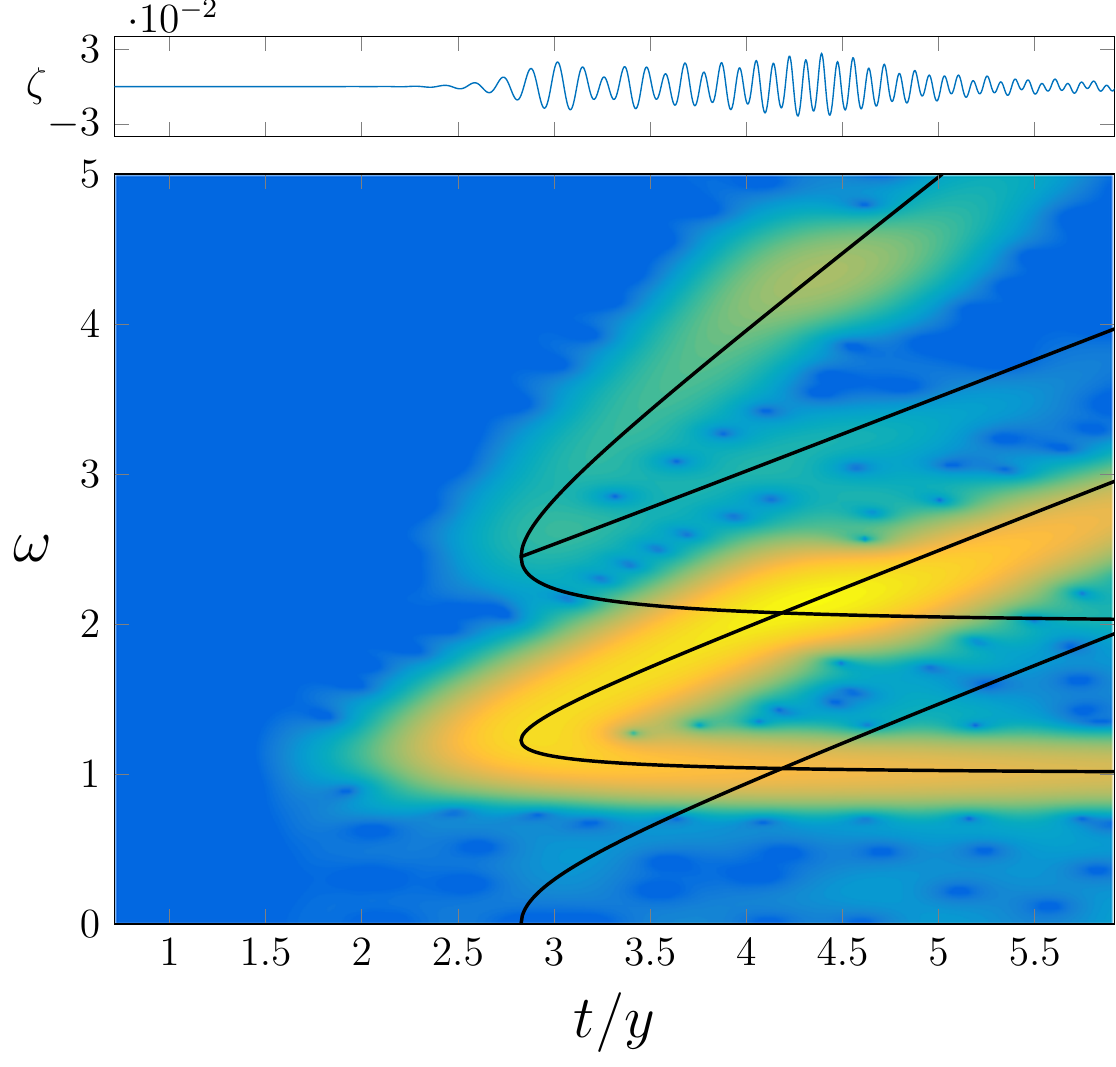}}&\\
\end{tabular}
\caption{Spectrograms of nonlinear ship waves for nondimensional pressure strength number $\epsilon=0.01$, $0.05$, $0.11$, $0.15$ and Froude number $F=1$. The solid curves and colour intensity are the same as in Figure~\ref{fig:nonlinspecFr02}.
}
\label{fig:nonlinspecFr10}
\end{figure}

{We now present in Figures~\ref{fig:nonlinspecFr02}-\ref{fig:nonlinspecFr10} spectrograms computed using our fully nonlinear numerical simulations described in \S\ref{sec:numscheme} together with our linear and weakly nonlinear predictions (\ref{eq:specCurve}) and (\ref{eq:specCurve2})-(\ref{eq:specCurve3}).  For all of the nonlinear spectrograms in these figures, we have fixed our $y$-coordinate to be $y=35$, which we have found to be sufficiently large to prevent unwanted blurring between branches of the dispersion curves.}

Figure \ref{fig:nonlinspecFr02} shows the spectrograms { computed} for Froude number $F=0.2$, representative of a slow moving `ship', and the four pressure strengths $\epsilon=0.15$, $0.75$, $1.65$ and $2.25$.  Before { proceeding to analyse} these { spectrograms}, we note two features that are not of particular concern in this study.  First, the low frequency intensity present in all spectrograms in Figure \ref{fig:nonlinspecFr02} (the horizontal band near $\omega=0$) is caused by the local { change in surface height} due to the { presence of the} pressure distribution, and is not part of the { far field wave train.  If we had the capacity to employ much larger computational domains for our numerical solution, then we could have fixed the $y$-coordinate to be larger than $y=35$, which would begin to eliminate this effect.  Either way, we ignore this low frequency band.} { Second, a signal is visible in the spectrogram as a band of intensity at $\omega=1$ ahead of the dispersion curve (for $t/y<\sqrt{8}$).  This effect is due to spurious numerical waves that are caused by numerical truncation {and the stitching method}, as discussed at the end of \S\ref{sec:numscheme}}.

Returning to the important trends in Figure \ref{fig:nonlinspecFr02}, we see that for low nonlinearity ($\epsilon=0.15$), the intensity in the spectrogram (Figure \ref{fig:nonlinspecFr02}(a)) occurs on the lower branch of the linear dispersion curve, $\omega_2$.  As the nonlinearity increases, the spectrogram intensity moves up the lower branch towards the fold of the linear dispersion curve (Figure \ref{fig:nonlinspecFr02}(b), $\epsilon=0.75$). These observations are consistent with our linear spectrograms in Figure~\ref{fig:linspec}, which show the dominant part of the time-frequency signal for slowly moving ships lies on the transverse wave component (the lower branch of the linear dispersion curve, $\omega_2$).  For even higher nonlinearity ($\epsilon=1.65$ in Figure~\ref{fig:nonlinspecFr02}(c) and $\epsilon=2.25$ in Figure~\ref{fig:nonlinspecFr02}(d)), additional high intensity portions appear along the second-order mode $\omega_4$, which comes from transverse waves interacting with themselves (recall $\omega_4=2\omega_2$).  There is even a hint of a further signal at $\omega=3\omega_2$, which would represent an even higher order mode.  Finally, the extremely nonlinear solution in Figure \ref{fig:nonlinspecFr02}(d) exhibits multiple modes appearing in a vertical section in line with the fold ($t/y=\sqrt{8}$).  These multiple modes bare a striking resemblance to the so-called leading waves shown in Figure~\ref{fig:torsvikspec}.  Thus we can conclude that the leading wave component identified by \citet{torsvik15a} is { possibly} due to nonlinearity (steep nonlinear waves).

For $F=0.7$, representative of a faster moving `ship', the maximum intensity of the spectrogram for the linear solution is around the fold of the linear dispersion curve (see Figure \ref{fig:linspec}(b)).  This behaviour is replicated in the nonlinear spectrograms in Figure \ref{fig:nonlinspecFr07}(a), which is for the small value $\epsilon=0.01$.  For moderate nonlinearity, shown in Figure \ref{fig:nonlinspecFr07}(b), an additional mode of intensity appears along $\omega_3$ and $\omega_5$.  Further increasing nonlinearity allows for a clearer realisation of the additional modes and the appearance of colour intensity along $\omega_6$ (Figure \ref{fig:nonlinspecFr07}(c)).  For a highly nonlinear solution, $\epsilon=0.15$ (Figure \ref{fig:nonlinspecFr07}(d)), the region of intensity distorts away from the linear dispersion curve, leading to high intensity regions of the time-frequency map appearing to the left of the fold (that is, for $t/y<\sqrt{8}$).  Thus, for high nonlinear flows past a pressure distribution, there exist a visible part of the wave train that appears outside of Kelvin's wedge, as discussed by \citet{pethiyagoda14b} (that is, some highly nonlinear solutions have apparent wake angles that are greater than Kelvin's angle).

Figure \ref{fig:nonlinspecFr10} is for $F=1$, representing an even faster `ship'.  Here the spectrogram follows the same trend as $F=0.7$ with a clearer colour intensity along the second-order dispersion curves $\omega_{3,5}$ and an absence of colour intensity along $\omega_6$. Thus, for highly nonlinear flows due to faster ships, we expect additional high intensity portions of the spectrogram along the second-order modes that come from divergent waves interacting with themselves ($\omega_3=2\omega_1$) and divergent waves interacting with transverse waves ($\omega_5=\omega_1+\omega_2$).  Further, we see that the distortion observed in Figure \ref{fig:nonlinspecFr10}(d) occurs later than that observed for $F=0.7$  (Figure \ref{fig:nonlinspecFr07}(d)). The different locations for the distortions is consistent with the location of the greatest intensity in the spectrograms for the exact linear solution.

We close this subsection by commenting on the ultimate fate of nonlinear solutions as our parameter $\epsilon$ increases.  From Bernoulli's equation (\ref{eq:bern}), we see there is an absolute upper bound for the free surface height, $\zeta_{\mathrm{upper}}=1/2$.  Thus, for a given solution, the maximum surface height $\zeta_{\mathrm{max}}<\zeta_{\mathrm{upper}}$.  This maximum surface height typically occurs at the crest of one of the divergent waves, although it can also occur on the centreline, depending on the parameter values.  As $\epsilon$ increases, the problem becomes more nonlinear, and we find the maximum surface height $\zeta_{\mathrm{max}}$ also increases.  Qualitatively, we observe that for increasing $\epsilon$, the waves themselves have sharper crests and broader troughs.  We speculate that ultimately there exists a critical value, $\epsilon_{\mathrm{c}}(F)$, such that $\zeta_{\mathrm{max}}\rightarrow\zeta_{\mathrm{upper}}$ as $\epsilon\rightarrow\epsilon_{\mathrm{c}}$, although our present numerical scheme is not capable of computing solutions that are sufficiently nonlinear to explore this issue in any detail.

\subsection{Comparison with experimental data}

{ We shall now make some comparisons with the experimental spectrogram shown in Figure \ref{fig:torsvikspec}.  The high speed ferry that produced the wake in question is named the \emph{Star}, which sails from Tallinn, Estonia to Helsinki, Finland.  The reported operating speed of the \emph{Star} at the time the data was measured was 14.2 ms$^{-1}$ \citep{parnell08}.  Its length and width are $186$ m and $27.7$ m, respectively.  This corresponds to a half-length based Froude number of $F_L\approx 0.47$ and half-width based Froude number of $F_W \approx 1.21$.  Obviously, the ferry has an aspect ratio which is much larger than unity.  However, it is still instructive to compare with our mathematical model which assumes an axisymmetric disturbance.}

\begin{figure}
\centering
\includegraphics[width=.6\linewidth]{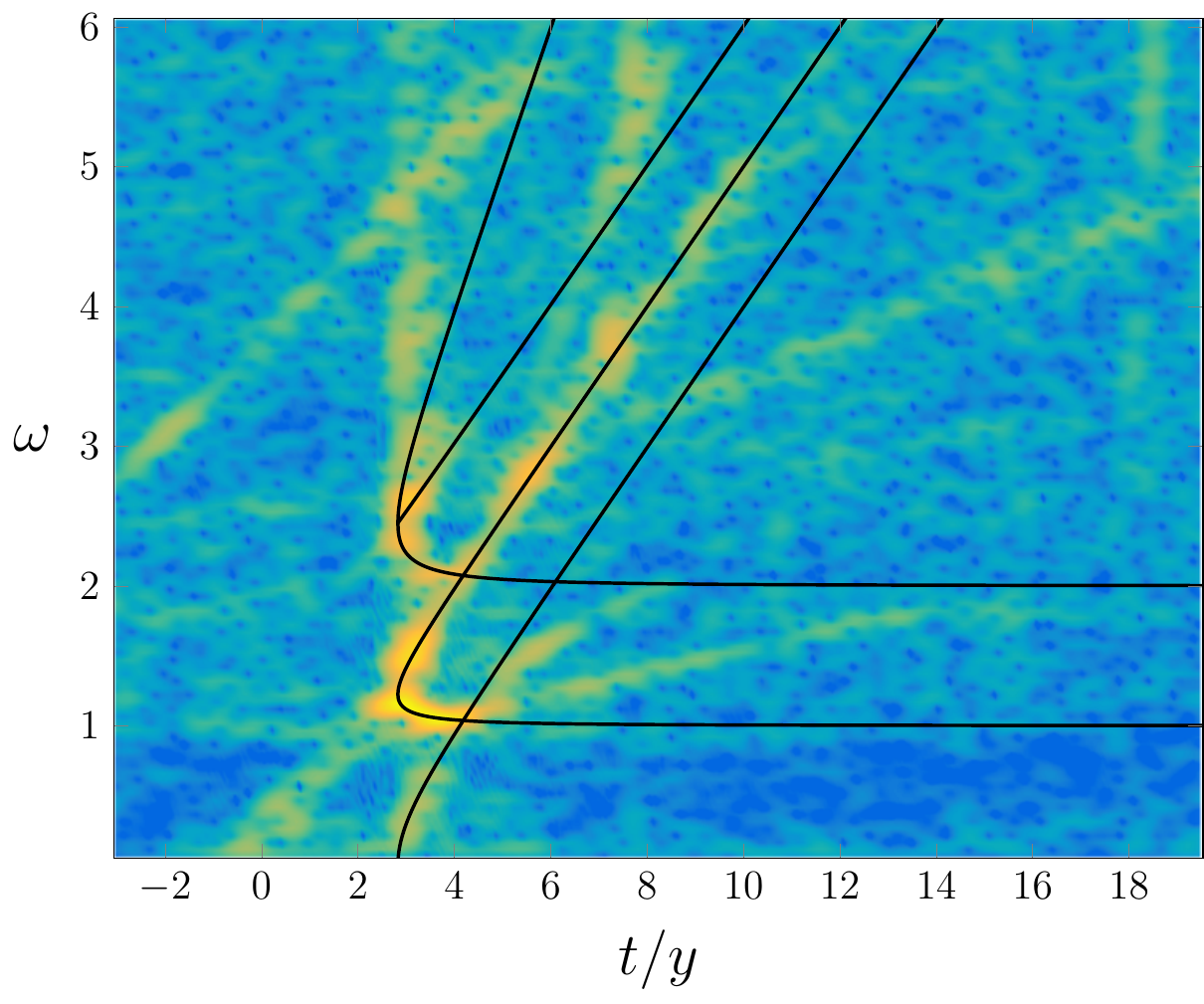}
\caption{A scaled version of Figure \ref{fig:torsvikspec}, with the dispersion curves given by (\ref{eq:specCurve}), (\ref{eq:specCurve2}) and (\ref{eq:specCurve3}) overlaid. The intensity of the spectrogram closely follows the linear and second order dispersion curves $\omega_{1-6}$ within the leading wave component of the spectrogram and follows the divergent wave portion of the linear dispersion curve $\omega_1$ outside the leading wave. The transverse wave component of the spectrogram does not appear to follow the lower branch of the linear dispersion curve, $\omega_2$, for large $t/y$.}
\label{fig:torsvikDispCurve}
\end{figure}

Considering the spectrogram in Figure \ref{fig:torsvikspec}, we now overlay the linear and second-order dispersion curves in order to compare our theoretical results with experimental data. To do so we scale the axes based on the speed of the ship, $U$, and the minimum distance to the sensor, $y$, and align the time the ship is closest to the sensor with $t=0$.  As the exact speed, distance and passing time are not known, we match the divergent wave intensity and fold location to the upper branch of the linear dispersion curve, $\omega_1$.  The resulting scaled spectrogram is shown in Figure~\ref{fig:torsvikDispCurve}.  Note that in this example we do not consider the transverse wave component $\omega_2$ when matching between our theoretical and experimental results, because the lower branch of the linear mode in the experimental spectrogram in Figure~\ref{fig:torsvikspec} does not appear to be horizontal for large $t/y$, as the theory predicts.  We return to this point in \S\ref{sec:acceleration}.

By matching the experimental spectrogram as just described, we find the speed of the ship to be 15.75 ms$^{-1}$, the distance from the sensor to be 2.5 km and the time the ship is closest to the sensor to be at 21:18. Comparing the properties of the ship's voyage with the {reported operating speed and the closest distance to the sensor given by \citet{torsvik15a}}, we see that our calculated speed is a slight over estimation of the reference value, 14.2 ms$^{-1}$.  This could be due to a number of issues, for example { the effects of a steady underlying current or} finite depth effects that subtly change the shape of the dispersion curve.  More encouragingly, having a clear divergent wave component means our estimated distance to the sensor falls in the reference bounds of 2.5--3 km, where 2.5 km corresponds to the outgoing shipping lane.

Visually, Figure~\ref{fig:torsvikDispCurve} shows very good agreement between the dispersion curves and the experimental spectrogram in the time frame of the leading wave.  In particular, for the region near $t/y=\sqrt{8}$ (roughly $2<t/y<4$), the high intensity part of the spectrogram follows the linear dispersion curve (made up of branches $\omega_{1,2}$) and the second-order curves (with branches $\omega_{3-6}$).  This exercise shows how important it is to understand the consequences of steep nonlinear waves when interpreting spectrograms.  It is worth noting that the experimental spectrograms presented by  \citet{wyatt88} (calculated from a tug boat named {\em Quapaw}) show very similar features, including the prominent high intensity regions in the leading wave component.  Indeed, \citet{wyatt88} comment that these properties are likely due to nonlinearity.

Finally, we note that to the right of the leading wave region of the spectrogram in Figure~\ref{fig:torsvikDispCurve} (for roughly $t/y>4$), there is an obvious disagreement between the experimental results and the lower (transverse) branch of the linear dispersion curve, $\omega_2$.  We consider this issue further now.

\section{An accelerating ship}\label{sec:acceleration}

\subsection{Linear dispersion curve with acceleration}

One feature of the experimental spectrogram in Figure \ref{fig:torsvikDispCurve} that is not yet explained is the obvious difference between the transverse branch of the linear dispersion curve, $\omega_2$, and the transverse component of the experimental spectrogram data.  In an attempt to explain this discrepancy, we consider the case that the ship is accelerating from rest up to its cruising speed.  We are motivated to explore this approach for two reasons: a ship travelling at a slower speed will generate transverse waves of higher frequency (from the dispersion relation); and the transverse waves are caused by waves generated earlier than those that generate divergent waves.

We illustrate the second property by referring to Figure \ref{fig:torsvikDispCurveAccel}(a) and (b).  In Figure~\ref{fig:torsvikDispCurveAccel}(a), the solid curve represents a hypothetical ship speed versus time for a $y$ value that is fixed to be the closest distance to the sensor.  The actual curve is found by fitting to data, as explained below.  The points A, B and C in the figure are associated with waves that are generated at a time $t_\mathrm{gen}<0$ (Figure \ref{fig:torsvikDispCurveAccel}(a)) and detected by the sensor at a later time $t_\mathrm{sen}>0$ (Figure \ref{fig:torsvikDispCurveAccel}(b)).  With this particular velocity profile, the wave represented by the point A in Figure \ref{fig:torsvikDispCurveAccel}(a)-(b), generated at $t_\mathrm{gen}\approx-6.05$ and detected by the sensor at $t_\mathrm{sen}\approx 18.34$, falls on the transverse branch of the dispersion curve.  The wave represented by the point B is generated later than A, at $t_\mathrm{gen}\approx-4.4$, but still lies on the transverse branch of the dispersion curve.  It is detected earlier by the sensor, at $t_\mathrm{sen}\approx 5.61$.  Waves generated at later times will eventually lie on the divergent wave branch, such as the wave represented by C.  Thus, from this argument we see that if the ship is travelling at a slower speed when wave A is generated compared to when wave B is generated, the frequency of the wave A along the dispersion curve will be greater than the frequency of wave B if both waves fall on the transverse branch of the dispersion curve. Therefore, the transverse branch of the linear dispersion curve will grow in frequency as $t/y$ increases if the ship is accelerating.

To determine the location of the linear dispersion curve for an accelerating ship, we first define the nondimensional displacement of the ship, $X(t)$,  and its velocity, $U(t)$, with the following properties:
\begin{equation*}
X(0)=0,\quad
U(t)=\frac{\mathrm{d}X}{\mathrm{d}t},\quad
0\leq\,U\leq 1,\quad
U(t)=1\,\,\mathrm{for}\,\,t>0
\end{equation*}
Here $t=0$ corresponds to the time at which the ship is closest to the sensor. Thus, under these properties, the ship is accelerating during an interval of time before $t=0$ but is moving with constant speed by the time $t=0$. The dispersion curve can be defined parametrically in terms of $\theta$ by first determining where the ship generated the wave, $X_\mathrm{gen}=-y\cot\theta$, and thus the time the wave was generated, $t_\mathrm{gen}\leq 0$ such that $X(t_\mathrm{gen})-X_\mathrm{gen}=0$. The time taken for the wave to reach the sensor can then be calculated by $t_\mathrm{wave}=y\,\mathrm{cosec}\,\theta/c_g$ where the group velocity $c_g=c_p/2=U(t_\mathrm{gen})\cos\theta/2$. Finally, the variable speed version of equations (\ref{eq:kFunc}) and (\ref{eq:omegaFunc}) given by $k(\theta)=\sec^2\theta/U^2$ and $\omega=Uk(\theta)\cos\theta$, respectively, are used to derive the dispersion curve
\begin{equation}
\left(\frac{t}{y},\omega\right)=\left(\frac{t_\mathrm{wave}+t_\mathrm{gen}}{y},\frac{\sec\theta}{U(t_\mathrm{gen})}\right). \label{eq:specCurveVarSpeed}
\end{equation}

\subsection{Comparison with experimental data}

\begin{figure}
\centering
\subfloat[Velocity function]{\includegraphics[width=.6\linewidth]{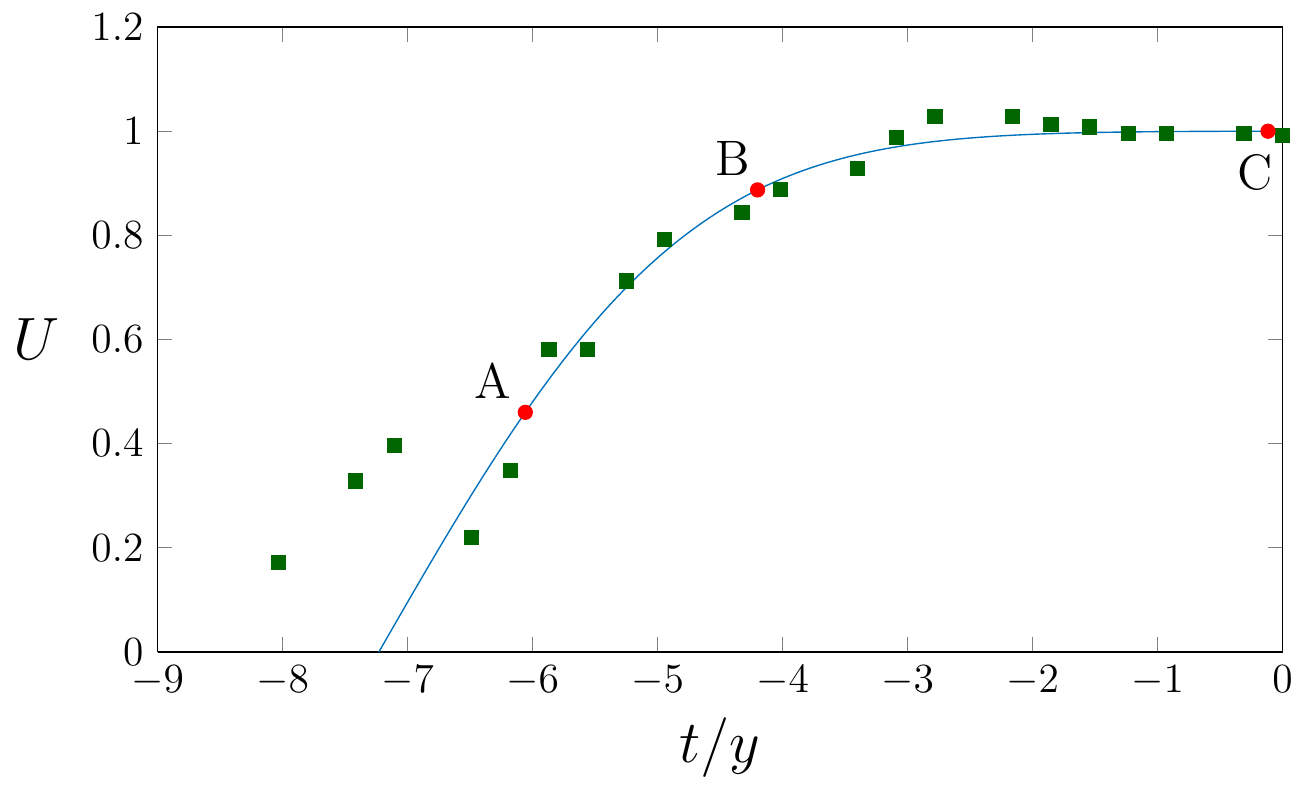}}\\
\subfloat[Linear dispersion curves]{\includegraphics[height=.4\linewidth]{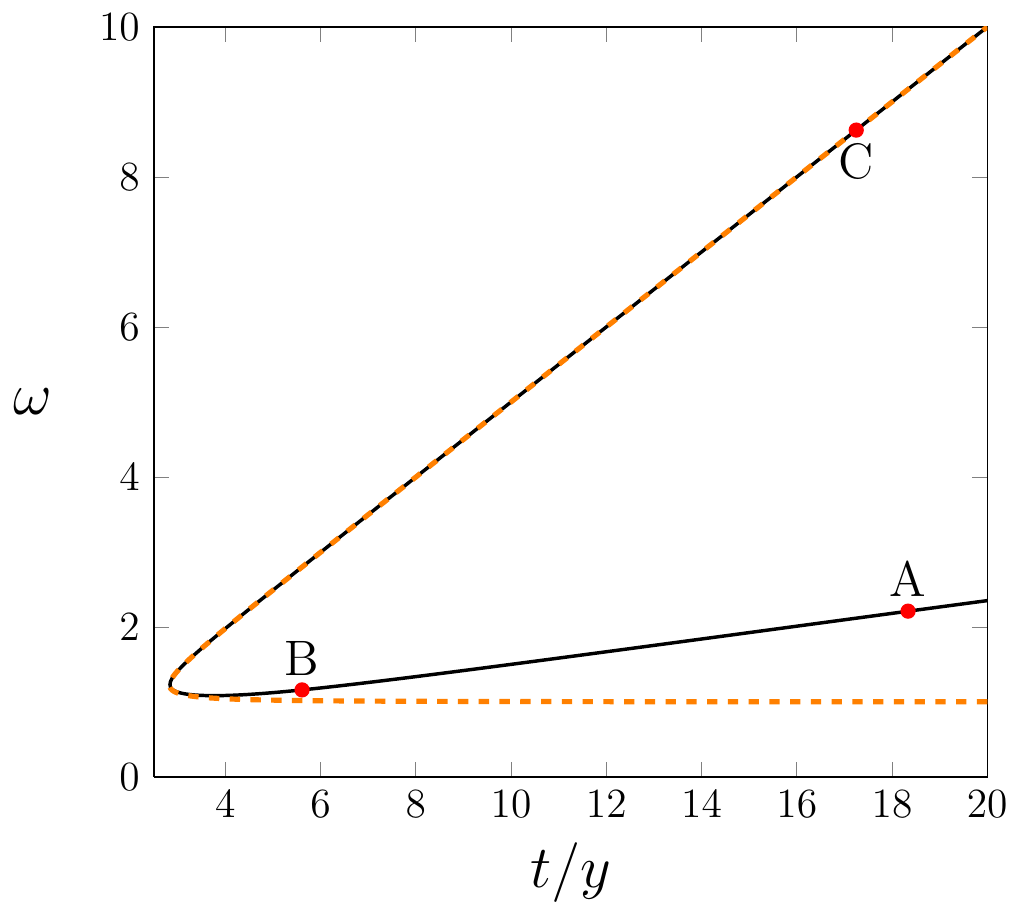}}
\subfloat[Experimental spectrogram]{\includegraphics[height=.4\linewidth]{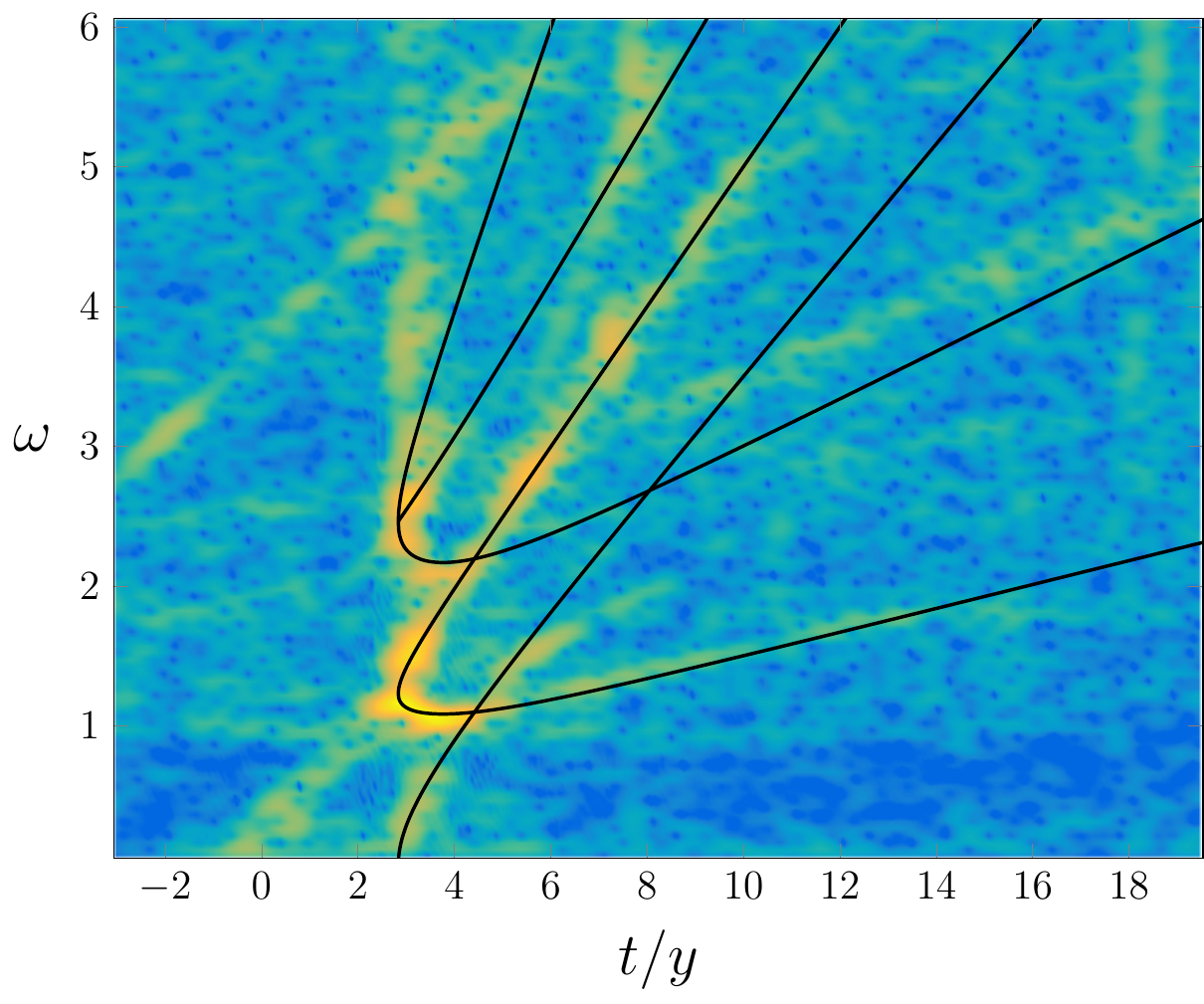}}
\caption{(a) A plot of example data (squares) for nondimensional ship speed and the roughly fitted velocity function (\ref{eq:velocity}). (b) A plot of the comparison between the linear dispersion curve for an accelerating ship (\ref{eq:specCurveVarSpeed}, solid) and constant-speed linear dispersion curve ($\omega_{1,2}$, dashed). (c) The dispersion curves computed for the accelerating ship overlaid on the scaled version of Figure \ref{fig:torsvikspec}.  The labels A, B and C in (a) and (b) each identify a wave generated at the ship at a time in (a) and received by the sensor at the time in (b).}
\label{fig:torsvikDispCurveAccel}
\end{figure}

The specific velocity function $U(t)$ will affect the shape of the new dispersion curve (\ref{eq:specCurveVarSpeed}), thus in order to match with the experimental spectrogram (Figure \ref{fig:torsvikspec}) we require a rough approximation for the velocity of the ship observed in the experiment.  We used the shipping traffic website run by the organisation MarineTraffic (MarineTraffic.com) to track the velocity of the \emph{Star} on a particular day as it left port at Tallinn and travelled to the point closest to the sensor (the location of the sensor is given by \citet{parnell08}). The data we obtained is nondimensionalised by scaling the velocity by the cruising speed, $U_\mathrm{cruise}$, and by scaling time by $U_\mathrm{cruise}/g$.  Time is shifted so that $t=0$ corresponds to when the ship was closest to the sensor. An example of the data (for a day in February 2016) is given as the squares in Figure \ref{fig:torsvikDispCurveAccel}(a).  It appears the ferry initially accelerates as it leaves port then slows down as it turns onto its sailing line before finally accelerating up to its cruising speed.  For this specific dataset $U_\mathrm{cruise}\approx 12.86\,\mathrm{ms}^{-1}$. Considering only the final acceleration phase, the velocity data for this particular ship looks roughly like an error function, thus we represent it by
\begin{equation}
U(t)=\mathrm{erf}\left(\frac{t/y-t_\mathrm{shift}}{\beta}\right),\label{eq:velocity}
\end{equation}
where $t_\mathrm{shift}=-535/74$ and $\beta=100/37$ are parameters that are determined by roughly fitting to the data.  The velocity profile (\ref{eq:velocity}) with these parameters is the solid curve in Figure \ref{fig:torsvikDispCurveAccel}(a).

Figure \ref{fig:torsvikDispCurveAccel}(b) shows a comparison between the new linear dispersion curve (\ref{eq:specCurveVarSpeed}) with $U(t)$ given by (\ref{eq:velocity}) for an accelerating ship (solid line) and the constant-speed linear dispersion curve ($\omega_{1,2}$, dashed). The transverse branch of the dispersion curve for the accelerating ship clearly increases in frequency as $t/y$ increases; conversely, the divergent branch is very close to the constant-speed dispersion branch. The second-order dispersion curves for the accelerating ship have been derived using the method in \S\ref{sec:nonlinear}, and the full set of dispersion curves has been overlaid onto the experimental spectrogram in Figure~\ref{fig:torsvikDispCurveAccel}(c), with the same axis-scaling as Figure~\ref{fig:torsvikDispCurve}. We see that the new linear dispersion curve does a much better job of predicting the transverse wave in the spectrogram.  Of course the fitting process used here is very rough, but our goal is only to suggest a reasonable approximation of the velocity $U(t)$ on a given day (the data for the ferry observed in 2008 is no longer freely available). Given the clear improvement in the comparison between the measured spectrogram and the theoretical prediction, we are confident that the increase in frequency in the measured transverse wave is due to the observed ship accelerating before passing by the sensor (and not due to nonlinearity, say).  { We acknowledge that other explanations are possible, for example shoaling or other finite-depth effects such as wave refraction.  Or perhaps another explanation entirely.}

\section{Discussion}
The use of spectrograms to analyse ship wakes has been explored recently in a number of studies \citep{benassai15,didenkulova13,sheremet13,torsvik15a,torsvik15b}. A key observation is that real-world spectrograms appear to highlight components of the wave signal that are in addition to the linear components associated with the traditional transverse and divergent wave systems.  Determining the cause of these additional components has provided the motivation for the present study.

Through the use of classical linear water wave theory and numerical simulations of nonlinear free-surface flow past a pressure distribution, we have identified both linear and higher-order modes present in the spectrograms.  As expected, the high intensity signal in the spectrogram for the linearised problem of flow past a pressure distribution follows the linear dispersion curve (\ref{eq:specCurve}).  By applying the analogy to flow past a ship, we have demonstrated that for slowly moving ships, the high intensity portion of the spectrogram lies on top of the transverse branch of the dispersion curve $\omega_2$ (also referred to as the constant-frequency mode), while for faster ships, the dominant part of the spectrogram lies on top of the divergent branch $\omega_1$ (the sliding-frequency mode).  We have applied a weakly nonlinear theory to calculate the location of second-order modes $\omega_{3-6}$ (see (\ref{eq:specCurve2})-(\ref{eq:specCurve3})); these do an excellent job of predicting high intensity regions in the spectrogram for a fully nonlinear numerical solutions.  This approach has allowed us to derive a quantitative description of the modes in the so-called leading wave component identified by \citet{torsvik15a}, including the parts that are due entirely to steep nonlinear waves, {without the need to include finite-depth effects in our modelling.}
Finally, we showed that the frequency increase in the measured transverse waves component of an experimental spectrogram is { possibly} due to the ship accelerating before passing by the sensor (and not due to nonlinearity). 

We are not in a position to provide an explanation for the cause of the so-called precursor solitary wave in in the spectrogram in Figure~\ref{fig:torsvikspec}, nor have we explained the existence of the two lines of intensity that appear to meet on $\omega_1$ at $t\approx 21:32$ and $f\approx 0.3$ in the same figure. There are many effects that could potentially be the cause of these additional features.  For example, the precursor solitary wave could be due to ``precursor solitons'' that are generated by vessels moving in shallow water \citep{soomere07}.  On the other hand, the two lines of intensity could be due to other slower vessels, either further away or closer to the sensor, or perhaps even a single accelerating vessel closer to the sensor.


The analytical results for the linear (\ref{eq:specCurve}) and second-order dispersion curves (\ref{eq:specCurve2})--(\ref{eq:specCurve3}) will hold for any problem in infinite depth for which the ship can be well approximated by a single disturbance.  These results could be easily extended to flows of constant finite depth by changing the dispersion relation, phase velocity and group velocity used in equations (\ref{eq:phaseVel})--(\ref{eq:thetaZero}).  The approach can also be extended to `ships' characterised by two disturbances (representing bow and stern waves) by including a second time shifted linear dispersion curve, and by considering up to four distinct waves (transverse and divergent waves from both disturbances) in the second order-approximation (\ref{eq:secOrder}).  Such an adjustment in the theory could be relevant for long thin ships \citep{noblesse14, zhu15}.  We note, however, that if the time shift between the two disturbances is sufficiently small, it will be difficult to differentiate between the two linear dispersion curves and, instead, the interference effects will be captured in the variation of the colour intensity.

Our work provides a deeper insight into the signature left behind in the wakes of ships.  This work is important because spectrogram analysis in the real world is only useful if users understand how to identify the key features of a time-frequency heat map and relate them to physical properties of the moving vessel.  As mentioned in the Introduction, an application of this work is to interpret data from an echo sounder placed in a shipping channel, with the goal of measuring the wave energy emitted by various vessels.  This information is important for monitoring damage to the coast, docked vessels or man-made structures.

\section*{Acknowledgements}
The authors wish to thank Tomas Torsvik for providing a copy of the experimental data in Figure~\ref{fig:torsvikspec} and Tarmo Soomere for some fruitful discussions.  SWM acknowledges the support of the Australian Research Council via the Discovery Project DP140100933. The authors thank Prof.\ Kevin Burrage for the use of high performance computing facilities and acknowledge further computational resources and support provided by the High Performance Computing and Research Support (HPC) group at Queensland University of Technology. Acknowledgement also goes to the three referees for their useful feedback.

\bibliographystyle{jfm}

\begin{thebibliography}{00}
\expandafter\ifx\csname natexlab\endcsname\relax\def\natexlab#1{#1}\fi

\bibitem[Benassai {\em et~al.\/}(2015)]{benassai15}
{\sc Benassai, G., Piscopo, V., \& Scamardella, A.} 2015 Spectral analysis of waves produced by HSC for coastal management. {\em J. Mar. Sci. Technol.}, 1--12.

\bibitem[Brown {\em et~al.\/}(1989)]{brown89}
{\sc Brown, E.~D., Buchsbaum, S.~B., Hall, R.~E., Penhune, J.~P., Schmitt, K.~F., Watson, K.~M., \& Wyatt, D.~C.} 1989 Observations of a nonlinear solitary wave packet in the Kelvin wake of a ship. {\em J. Fluid Mech.} {\bf 204}, 263--293.

\bibitem[Chung \& Lim(2013)]{chung91}
{\sc Chung, Y.~K. \& Lim, J.~S.} 1991 A review of the Kelvin ship wave pattern. {\em J. Ship Res.} {\bf 35}, 191--197.

\bibitem[Cohen(1989)]{cohen89}
{\sc Cohen, L.} 1989 Time-frequency distributions -- a review. {\em Proc. IEEE} {\bf 77}, 941--981.

\bibitem[Darmon {\em et~al.\/}(2014)]{darmon14}
{\sc Darmon, A., Benzaquen, M. \& Rapha\"{e}l, E.} 2014 Kelvin wake pattern at large Froude numbers. {\em J. Fluid Mech.} {\bf 738}, R3.

\bibitem[Darrigol(2003)]{darrigol03}
{\sc Darrigol, O.} 2003 The spirited horse, the engineer, and the
  mathematician: water waves in nineteenth-century hydrodynamics. {\em Arch.
  Hist. Exact Sci.\/} {\bf 58}, 21--95.

\bibitem[Didenkulova {\em et~al.\/}(2013)]{didenkulova13}
{\sc Didenkulova, I., Sheremet, A., Torsvik, T. \& Soomere, T.} 2013 Characteristic properties of different vessel wake signals. {\em J. Coast. Res.} {\bf SI 65}, 213--218.

\bibitem[Ellingsen(2014)]{ellingsen14}
{\sc Ellingsen, S.~\r{A}.} 2014 Ship waves in the presence of uniform
  vorticity. {\em J. Fluid Mech.\/} {\bf 742}, R2.

\bibitem[Forbes(1989)]{forbes89}
{\sc Forbes, L.~K.} 1989 An algorithm for 3-dimensional free-surface problems in hydrodynamics. {\em J. Comp. Phys.\/} {\bf 82}, 330--347.

\bibitem[Harris(1978)]{harris78}
{\sc Harris, F. J.} 1978 On the use of windows for harmonic analysis with the discrete Fourier transform. {\em Proc. IEEE} {\bf 66}, 51--83.

\bibitem[Havelock(1932)]{havelock32}
{\sc Havelock, T. H.} 1932 The theory of wave resistance. {\em Proc. Roy. Soc. Lond.} A {\bf 138}, 339--348.


\bibitem[Hogben(1972)]{hogben72}
{\sc Hogben, N.} 1972 Nonlinear distortion of the Kelvin ship-wave pattern. {\em J. Fluid Mech.} {\bf 55}, 513--528.

\bibitem[Kurennoy {\em et~al.\/}(2009)Kurennoy, \textbf{Scully-Power} \& Parnell]{kurennoy09}
{\sc Kurennoy, D., Soomere, T. \& Parnell, K.} 2009 Variability in the properties of wakes generated by high-speed ferries. {\em J. Coast. Res.} {\bf 56},  519--523.



\bibitem[Rabaud \& Moisy(2013)]{rabaud13}
{\sc Rabaud, M. \& Moisy, F.} 2013 Ship wakes: {K}elvin or {M}ach angle? {\em
  Phys. Rev. Lett.\/} {\bf 110}, 214503.

\bibitem[Li \& Ellingsen(2016)]{li16}
{\sc Li, Y. \& Ellingsen, S.~\r{A}.} 2016 Ship waves on uniform shear current at finite depth: wave resistance and critical velocity {\em J. Fluid Mech.} {\bf 791}, 539--567.

\bibitem[MarineTraffic]{marinetraffic}
{\sc MarineTraffic} AIS Vessel Tracking - AIS Positions Maps. Retrieved 26 February 2016, from http://www.marinetraffic.com/en/ais/home/shipid:352956/zoom:10

\bibitem[Maruo(1967)]{maruo67}
{\sc Maruo, H.} 1967 High-and low-aspect ratio approximation of planing surfaces. {\em Schiffstechnik} {\bf 14}, 57--64.


\bibitem[Michell(1898)]{michell98}
{\sc Michell, J.~H.} 1898 The wave-resistance of a ship. {\em Phil. Mag.} {\bf 45}, 106--123.

\bibitem[Milgram(1988)]{milgram88}
{\sc Milgram, J. H.} 1988 Theory of radar backscatter from short waves generated by ships, with application to radar (SAR) imagery. {\em J. Ship Res.} {\bf 32}, 54--69.


\bibitem[Munk {\em et~al.\/}(1986)Munk, Scully-Power \& Zachariasen]{munk86}
{\sc Munk, W.~H., Scully-Power, P. \& Zachariasen F.} 1986 The Bakerian lecture, 1986. Ships from space. {\em Proc. Roy. Soc. London. A.} {\bf 412}, 231--254.

\bibitem[Noblesse(1981)]{noblesse81}
{\sc Noblesse, F.} 1981 Alternative integral representations for the Green function of the theory of ship wave resistance.  {\em J. Eng. Math.} {\bf 15}, 241--265.

\bibitem[Noblesse {\em et~al.\/}(2014)Noblesse, He, Zhu, Hong, Zhang, Zhu \&
  Yang]{noblesse14}
{\sc Noblesse, F., He, J., Zhu, Y., Hong, L., Zhang, C., Zhu, R. \& Yang, C.}
  2014 Why can ship wakes appear narrower than {Kelvin's} angle? {\em Euro. J.
  Mech. B/Fluids\/} {\bf 46}, 164--171.

\bibitem[Rabaud \& Moisy(2013)]{rabaud13}
{\sc Rabaud, M. \& Moisy, F.} 2013 Ship wakes: {K}elvin or {M}ach angle? {\em
  Phys. Rev. Lett.\/} {\bf 110}, 214503.

\bibitem[P\u{a}r\u{a}u \& Vanden-Broeck(2002)]{parau02}
{\sc P\u{a}r\u{a}u, E. \& Vanden-Broeck, J.-M.} 2002 Nonlinear two- and
  three-dimensional free surface flows due to moving disturbances. {\em Euro.
  J. Mech. B/Fluids\/} {\bf 21}, 643--656.

\bibitem[P\u{a}r\u{a}u {\em et~al.\/}(2007)P\u{a}r\u{a}u, Vanden-Broeck \& Cooker]{parau07}
{\sc P\u{a}r\u{a}u, E., Vanden-Broeck, J.-M. \& Cooker, M.~J.}
2007 Nonlinear three-dimensional interfacial flows with a free surface.  {\em J. Fluid Mech.} {\bf 591}, 481--494.

\bibitem[Parnell {\em et~al.\/}(2008)Parnell, Delpeche, Didenkulova, Dolphin, Erm, Kask, Kelp\v{s}aite, Kurennoy, Quak \& R\"{a}\"{a}met]{parnell08}
{\sc Parnell, K., Delpeche, N., Didenkulova, I., Dolphin, T., Erm, A., Kask, A., Kelp\v{s}aite, L., Kurennoy, D., Quak, E. \& R\"{a}\"{a}met, A.} 2008. Far-field vessel wakes in Tallinn Bay. {\em Est. J. Eng.} {\bf 14}, 273--302.


\bibitem[Peters(1949)]{peters49}
{\sc Peters, A.~S.} 1949 A new treatment of the ship wave problem. {\em Comm. Pure Appl. Math.} {\bf 2}, 123--148.

\bibitem[Pethiyagoda {\em et~al.\/}(2014a)Pethiyagoda, McCue, Moroney \& Back]{pethiyagoda14a}
{\sc Pethiyagoda, R., McCue, S.~W., Moroney, T.~J. \& Back, J.~M.} 2014a
  Jacobian-free {N}ewton-{K}rylov methods with {GPU} acceleration for computing
  nonlinear ship wave patterns. {\em J. Comp. Phys.\/} {\bf 269}, 297--313.

\bibitem[Pethiyagoda {\em et~al.\/}(2014b)Pethiyagoda, McCue \& Moroney]{pethiyagoda14b}
{\sc Pethiyagoda, R., McCue, S.~W. \& Moroney, T. J.} 2014b What is the apparent angle of a Kelvin ship wave pattern? {\em J. Fluid Mech.} {\bf 758}, 468--485.

\bibitem[Pethiyagoda {\em et~al.\/}(2015)Pethiyagoda, McCue \& Moroney]{pethiyagoda15}
{\sc Pethiyagoda, R., McCue, S.~W. \& Moroney, T. J.} 2015 Wake angle for surface gravity waves on a finite depth fluid {\em Phys. Fluids} {\bf 27}, 061701.

\bibitem[Reed \& Milgram(2002)Reed \& Milgram]{reed02}
{\sc Reed, A. M., \& Milgram, J. H.} 2002 Ship wakes and their radar images. {\em Ann. Rev. Fluid Mech.} {\bf 34}, 469--502.

\bibitem[Sheremet {\em et~al.\/}(2013)Sheremet, Gravois, U. \&  Tian]{sheremet13}
{\sc Sheremet, A.,Gravois, U. \&  Tian, M.} 2013 Boat-wake statistics at Jensen Beach, Florida. {\em J. Waterw. Port Coast. Ocean Eng.} {\bf 139}, 286--294.

\bibitem[Soomere(2007)]{soomere07}
{\sc Soomere, T.} 2007 Nonlinear components of ship wake waves. {\em Appl. Mech. Rev.} {\bf 60}, 120--138.

\bibitem[Thomson(1887)]{kelvin87}
{\sc Thomson, W.} 1887 On ship waves. {\em Proc. Inst. Mech. Engrs\/} {\bf 38},
  409--434.

\bibitem[Torsvik {\em et~al.\/}(2015a)Torsvik, Soomere, Didenkulova, Sheremet]{torsvik15a}
{\sc Torsvik, T., Soomere, T., Didenkulova, I. \& Sheremet, A.} 2015a Identification of ship wake structures by a time-frequency method. {\em J. Fluid Mech.\/} {\bf 765}, 229--251.

\bibitem[Torsvik {\em et~al.\/}(2015b)Torsvik, Herrmann, Didenkulova, Rodin]{torsvik15b}
{\sc Torsvik, T., Herrmann, H., Didenkulova, I. \& Rodin, A.} 2015b Analysis of ship wake transformation in the coastal zone using time-frequency methods. {\em Proc. Est. Acad. Sci.\/} {\bf 64}, 379--388.

\bibitem[Tuck(1975)]{tuck75}
{\sc Tuck, E. O.} 1975 Low-aspect-ratio flat-ship theory. {\em J. Hydronaut.} {\bf 9}, 3--12.


\bibitem[Tuck {\em et~al.\/}(1971)Tuck, Collins, Wells]{tuck71}
{\sc Tuck, E. O., Collins, J. I. \& Wells, W. H.} 1971 On ship wave patterns and their spectra. {\em J. Ship Res.} {\bf 15}, 11--21.


\bibitem[Ursell(1960)]{ursell60}
{\sc Ursell, F.} 1960 On Kelvin's ship-wave pattern. {\em J. Fluid Mech.} {\bf 8}, 418--431.

\bibitem[Wehausen \& Laitone (1960)]{wehausen60}
{\sc Wehausen, J.~V. \& Laitone, E.~V.} 1960 {\em Surface waves.} Springer.

\bibitem[Wyatt \& Hall(1988)]{wyatt88}
{\sc Wyatt, D.~C. \& Hall, R.~E.} 1988 Analysis of ship-generated surface waves using a method based upon the local Fourier transform. {\em J. Geophys. Res.} {\bf 93} (C11), 14133--14164.

\bibitem[Zhu {\em et~al.\/}(2015)Zhu, He, Zhang, Wu, Wan, Zhu, \& Noblesse]{zhu15}
{\sc Zhu, Y., He, J., Zhang, C., Wu, H., Wan, D., Zhu, R., \& Noblesse, F.} 2015 Farfield waves created by a monohull ship in shallow water. {\em Euro. J. Mech. B/Fluids} {\bf 49}, 226--234.
\end{thebibliography}

\end{document}